\documentclass[10pt,twocolumn,letterpaper]{article}

\usepackage{iccv}
\usepackage{times}
\usepackage{epsfig}
\usepackage{graphicx}
\usepackage{amsmath}
\usepackage{amssymb}
\usepackage{fancyhdr}
\usepackage{bm}
\usepackage{bbm}
\usepackage{mathtools}
\usepackage{xcolor}
\usepackage{soul}
\usepackage{booktabs, multirow}
\usepackage{enumitem}
\usepackage{subcaption}
\usepackage{xspace}
\usepackage{standalone}
\usepackage{balance}
\usepackage{scalerel}
\usepackage{selectp} 
\usepackage[ruled,vlined]{algorithm2e}

\usepackage[pagebackref=true,breaklinks=true,letterpaper=true,colorlinks,bookmarks=false]{hyperref}

\def\eg{{\em e.g.,}}
\def\ie{{\em i.e.,}}
\def\vs{{\em vs.}}
\def\etc{{\em etc}\xspace}
\newcommand{\one}[1]{\mathbbm{1}_{[#1]}}
\newcommand{\multisim}[0]{MICLe\xspace}

\newcommand{\figref}[1]{Fig.~\ref{#1}}

\SetKwInput{KwInput}{Input}                
\SetKwInput{KwOutput}{Output}              

\iccvfinalcopy 
\def\iccvPaperID{5551} 

\ificcvfinal\fi
\begin{document}

\title{Big Self-Supervised Models Advance Medical Image Classifications\\[-.1cm]}

\author{
Shekoofeh Azizi, 
Basil Mustafa,
Fiona Ryan\thanks{Former intern at Google. Currently at Georgia Institute of Technology.}\:, 
Zachary Beaver, 
Jan Freyberg, 
Jonathan Deaton,\\  
Aaron Loh, 
Alan Karthikesalingam, 
Simon Kornblith, 
Ting Chen, 
Vivek Natarajan, 
Mohammad Norouzi\\[.2cm]
Google Research and Health\thanks{{\{shekazizi,\,skornblith,\,iamtingchen,\,natviv,\,mnorouzi\}@google.com}}\:
}

\maketitle

\ificcvfinal\thispagestyle{empty}\fi

\vspace{-2pt}
\begin{abstract}
\vspace{-2pt}
Self-supervised pretraining followed by supervised fine-tuning has seen success in image recognition, especially when labeled examples are scarce, but has received limited attention in medical image analysis. This paper studies the effectiveness of self-supervised learning as a pretraining strategy for medical image classification.
We conduct experiments on two distinct tasks: dermatology condition classification from digital camera images and multi-label chest X-ray classification,
and demonstrate that self-supervised learning on ImageNet, followed by additional self-supervised learning on unlabeled domain-specific medical images
significantly improves the accuracy of medical image classifiers.
We introduce a novel Multi-Instance Contrastive Learning (MICLe) method that uses multiple images of the underlying pathology per patient case, when available, to construct more informative positive pairs for self-supervised learning.
Combining our contributions, we achieve an improvement of 6.7\% in top-1 accuracy and an improvement of 1.1\% in mean AUC on dermatology and chest X-ray classification respectively, outperforming strong supervised baselines pretrained on ImageNet. In addition, we show that big self-supervised models are robust to distribution shift and can learn efficiently with a small number of labeled medical images. 

\vspace{-2pt}

\end{abstract}

\setlength{\textfloatsep}{8pt}

\vspace{-2pt}
\section{Introduction}
\vspace{-2pt}

Learning from limited labeled data is a fundamental problem in machine learning, which is crucial for medical image analysis because annotating medical images is time-consuming and expensive. Two common pretraining approaches to learning from limited labeled data include: (1)~{\em supervised pretraining} on a large labeled dataset such as ImageNet, (2)~{\em self-supervised pretraining}  using contrastive learning (\eg~\cite{ he2019momentum, chen2020simple, chen2020big}) on unlabeled data.
After pretraining, supervised fine-tuning on a target labeled dataset of interest is used. While ImageNet pretraining is ubiquitous in medical image analysis~\cite{xie2019dual, menegola2017knowledge, mckinney2020international, liu2020deep, graziani2019visualizing, heker2020joint}, the use of self-supervised approaches has received limited attention. Self-supervised approaches are attractive because they enable the use of {\em unlabeled} domain-specific images during pretraining to learn more relevant representations.

\begin{figure}
    \centering
    \includegraphics[width=0.43\textwidth]{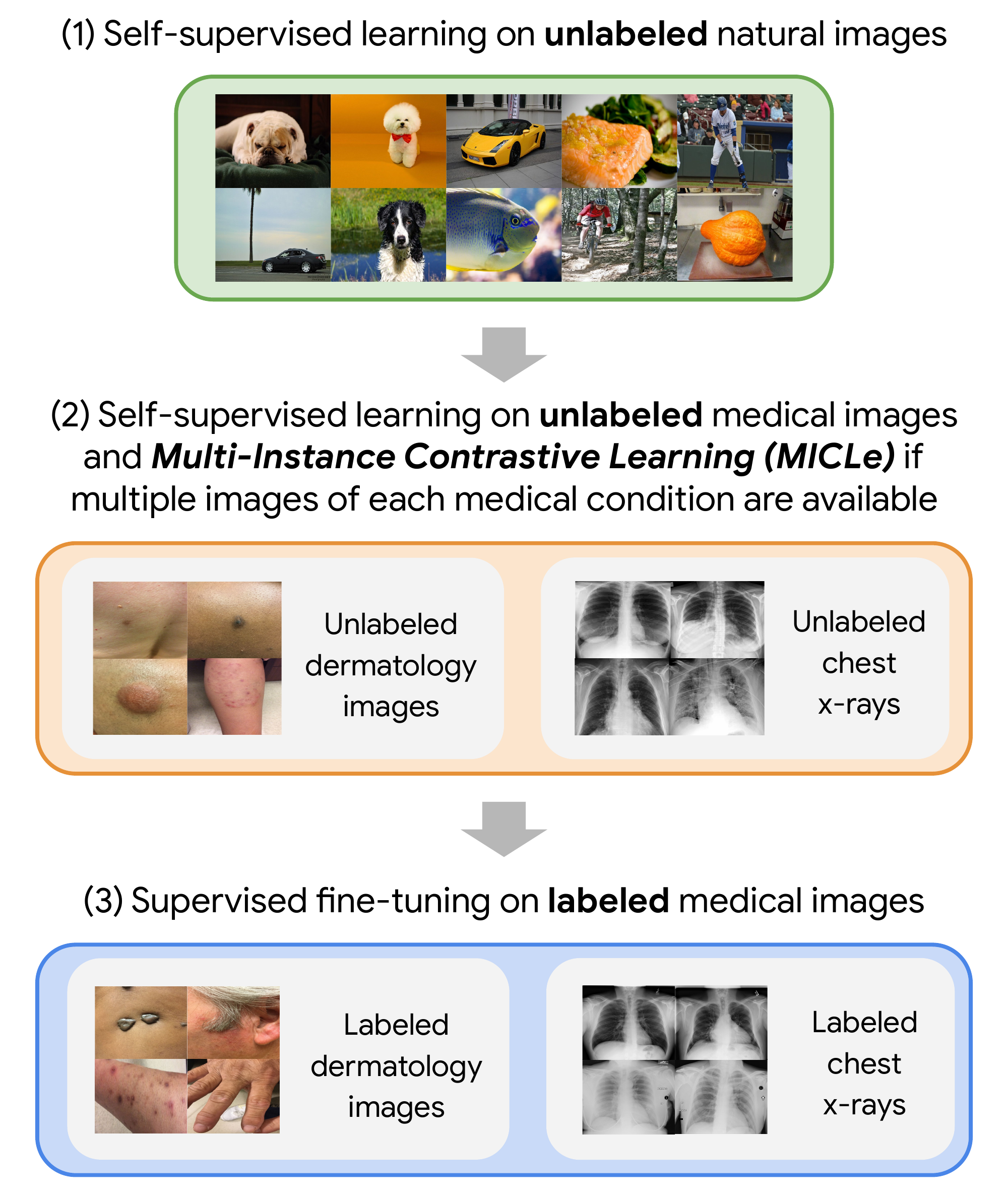}
    \vspace{-.3cm}
    \caption{\small
    Our approach comprises three steps: (1) Self-supervised pretraining on unlabeled ImageNet using SimCLR~\cite{chen2020simple}.
    (2) Additional self-supervised pretraining using unlabeled medical images. If multiple images of each medical condition are available,
    a novel Multi-Instance Contrastive Learning (MICLe) is used to construct more informative positive pairs based on different images.
    (3) Supervised fine-tuning on labeled medical images. Note that unlike step (1), steps (2) and (3) are task and dataset specific.}
    \label{fig:fig-1}
\end{figure}

\begin{figure}
     \centering
     \begin{subfigure}[b]{0.236\textwidth}
         \centering
         \includegraphics[width=\textwidth]{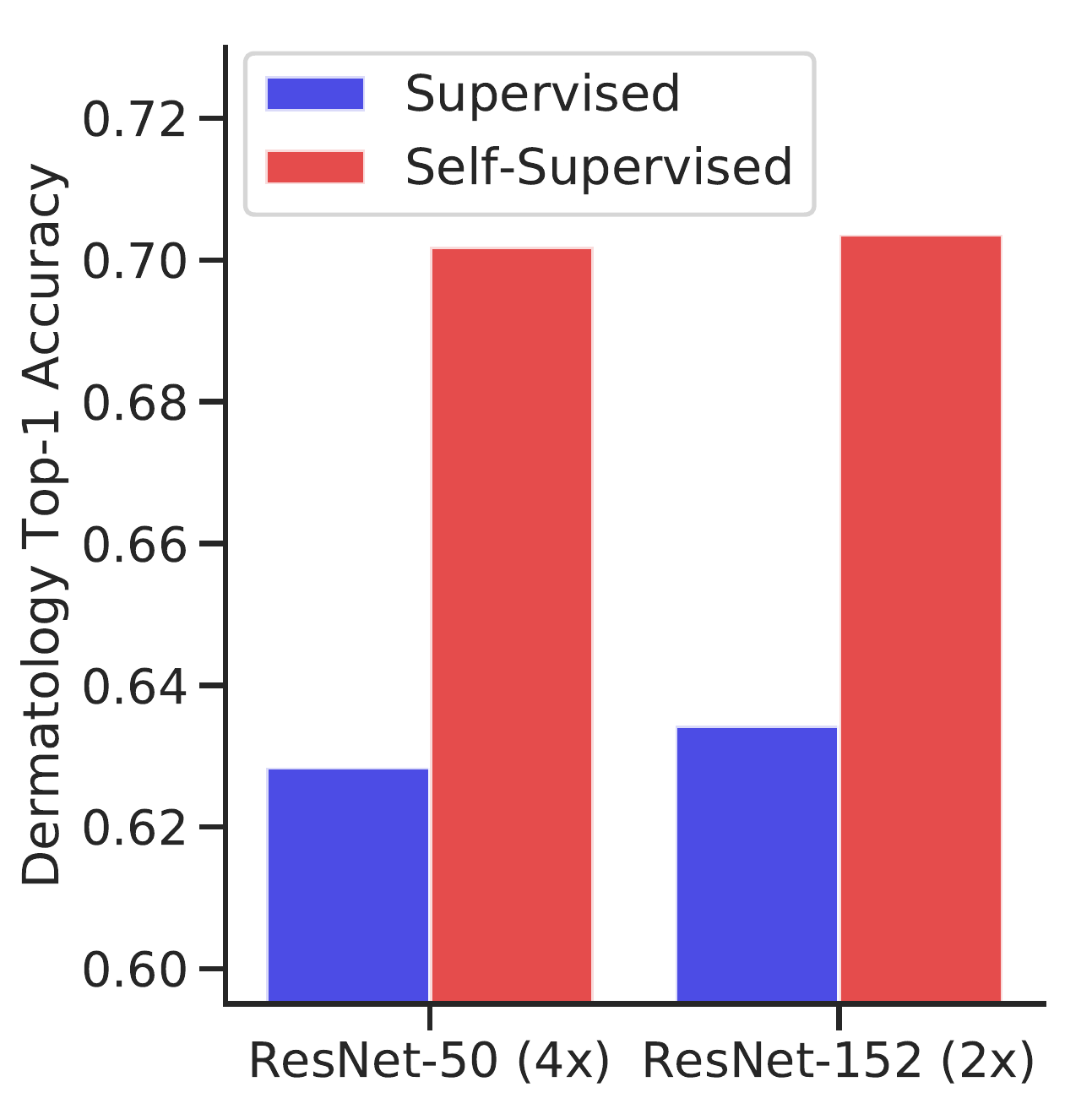}
     \end{subfigure}
     \begin{subfigure}[b]{0.236\textwidth}
         \centering
         \includegraphics[width=\textwidth]{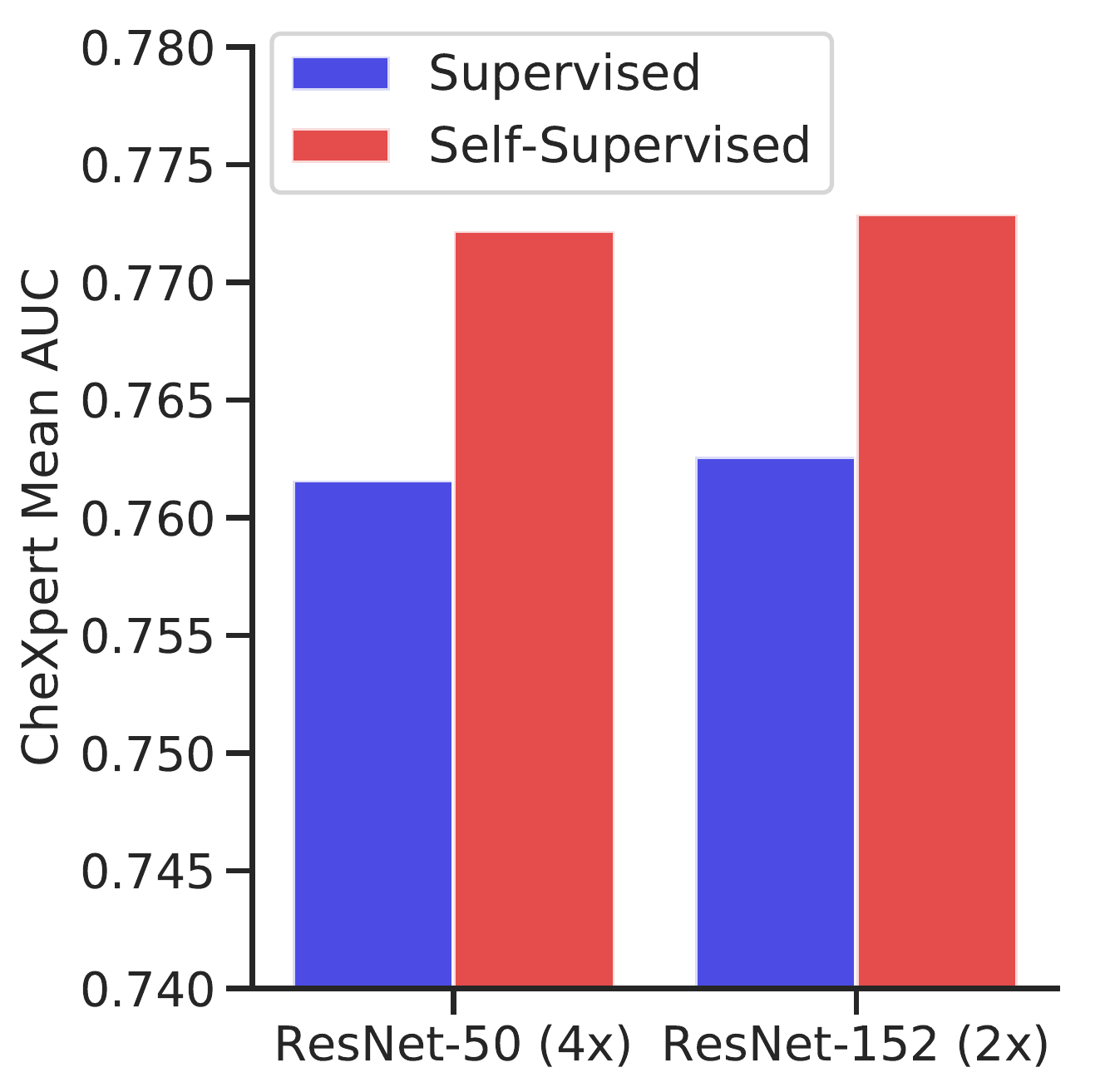}
     \end{subfigure}
        \vspace{-18pt}
        \caption{\small Comparison of supervised and self-supervised pretraining, followed by supervised fine-tuning using two architectures on dermatology and chest X-ray classification. Self-supervised learning utilizes unlabeled domain-specific medical images and significantly outperforms supervised ImageNet pretraining.}
        \label{fig:fig-2}
\end{figure}

This paper studies self-supervised learning for medical image analysis and conducts a fair comparison between self-supervised and supervised pretraining on two distinct medical image classification tasks: (1) dermatology skin condition classification from digital camera images, (2)~multi-label chest X-ray classification among five pathologies based on the CheXpert dataset~\cite{irvin2019chexpert}.
We observe that self-supervised pretraining outperforms supervised pretraining, even when the full ImageNet dataset (14M images and 21.8K classes) is used for supervised pretraining. We attribute this finding to the domain shift and discrepancy between the nature of recognition tasks in ImageNet and medical image classification. Self-supervised approaches bridge this domain gap by leveraging in-domain medical data for pretraining and they also scale gracefully as they do not require any form of class label annotation.

An important component of our self-supervised learning framework is an effective {\em Multi-Instance Contrastive Learning (\multisim)} strategy that helps adapt contrastive learning to multiple images of the underlying pathology per patient case. Such multi-instance data is often available in medical imaging datasets -- \eg~frontal and lateral views of mammograms, retinal fundus images from each eye, \etc.
Given multiple images of a given patient case, we propose to construct a positive pair for self-supervised contrastive learning by drawing two crops from two distinct images of the same patient case. Such images may be taken from different viewing angles and show different body parts with the same underlying pathology. This presents a great opportunity for self-supervised learning algorithms to learn representations that are robust to changes of viewpoint, imaging conditions, and other confounding factors in a direct way.
\multisim does not require class label information and only relies on different images of an underlying pathology, the type of which may be unknown.

\figref{fig:fig-1} depicts the proposed self-supervised learning approach, and \figref{fig:fig-2} shows the summary of results. Our key findings and contributions include:
\begin{itemize}[noitemsep,topsep=0pt,leftmargin=12pt]

\item We investigate the use of self-supervised pretraining on medical image classification. We find that self-supervised pretraining on unlabeled medical images significantly outperforms standard ImageNet pretraining and random initialization. 


\item We propose Multi-Instance Contrastive Learning (MICLe) as a generalization of existing contrastive learning approaches to leverage 
multiple images per medical condition. We find that \multisim improves the performance of self-supervised models, yielding state-of-the-art results.

\item On dermatology condition classification, our self-supervised approach provides a sizable gain of 6.7\% in top-1 accuracy, even in a highly competitive production setting.
On chest X-ray classification, self-supervised learning outperforms strong supervised baselines pretrained on ImageNet by 1.1\% in mean AUC.


\item We demonstrate that self-supervised models are robust and generalize better than baseslines, when subjected to shifted test sets, without fine-tuning. Such behavior is desirable for deployment in a real-world clinical setting.

\end{itemize}

\begin{figure*}[t]
\small
    \centering
    \includegraphics[width=1.0\textwidth]{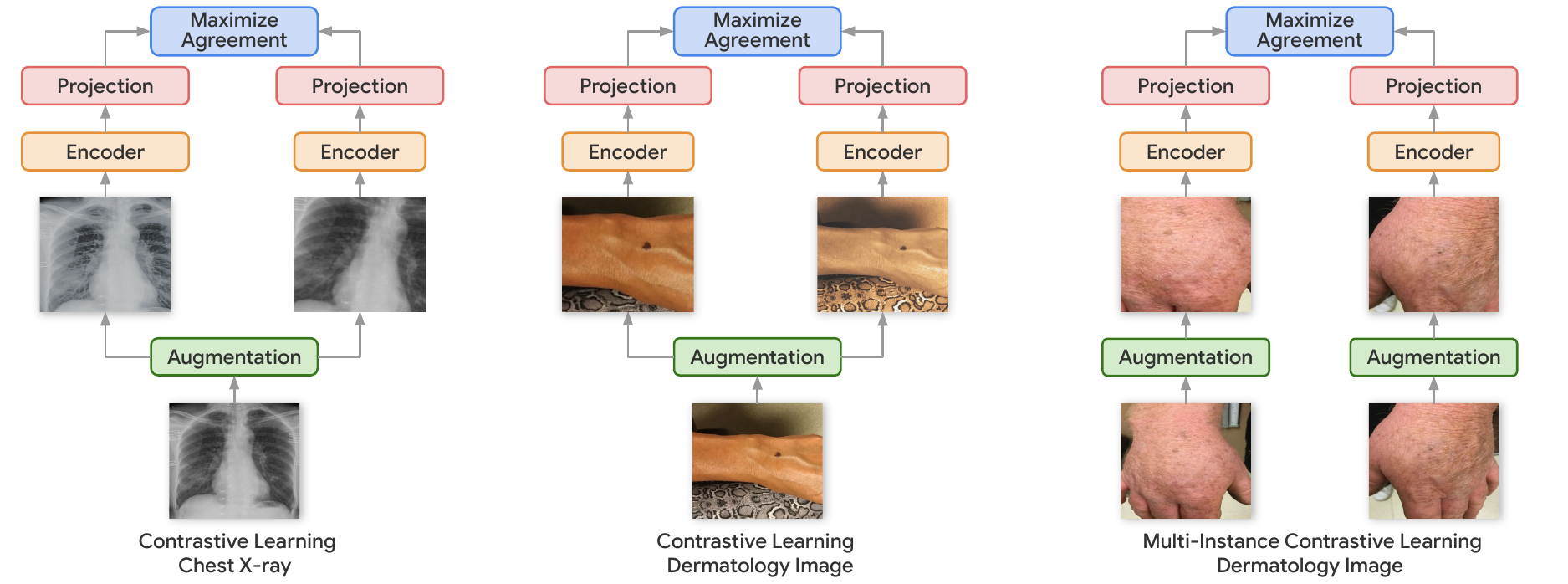}
    \vspace{-18pt}
    \caption{An illustrations of our self-supervised pretraining for medical image analysis. When a single image of a medical condition is available, we use standard data augmentation to generate two augmented views of the same image. When multiple images are available, we use two distinct images to directly create a positive pair of examples. We call the latter approach Multi-Instance Contrastive Learning (MICLe).}
    \vspace{-12pt}
    \label{fig:micle}
\end{figure*}

\vspace{-1pt}
\section{Related Work}
\vspace{-1pt}
\noindent {\bf Transfer Learning for Medical Image Analysis. } Despite the differences in image statistics, scale, and task-relevant features, transfer learning from natural images is commonly used in medical image analysis~\cite{liu2020deep,mckinney2020international,menegola2017knowledge,xie2019dual}, and multiple empirical studies show that this improves performance~\cite{alzubaidi2020towards, graziani2019visualizing, heker2020joint}. However, detailed investigations from Raghu \textit{et al.}~\cite{raghu2019transfusion} of this strategy indicate this does not always improve performance in medical imaging contexts. They, however, do show that transfer learning from ImageNet can speed up convergence, and is particularly helpful when the medical image training data is limited. Importantly, the study used relatively small architectures, and found pronounced improvements with small amounts of data especially when using their largest architecture of ResNet-50 (1$\times$)~\cite{he2016deep}. Transfer learning from in-domain data can help alleviate the domain mismatch issue. For example,~\cite{chen2019med3d,heker2020joint,liang2020transfer,pmlr-v102-geyer19a} report performance improvements when pretraining on labeled data in the same domain. However, this approach is often infeasible for many medical tasks in which labeled data is expensive and time-consuming to obtain. Recent advances in self-supervised learning provide a promising alternative enabling the use of unlabeled medical data that is often easier to procure.

\vspace{3pt}
\noindent {\bf Self-supervised Learning.} Initial work in self-supervised representation learning focused on the problem of learning embeddings without labels such that a low-capacity (commonly linear) classifier operating on these embeddings could achieve high classification accuracy~ \cite{doersch2015unsupervised,gidaris2018unsupervised,noroozi2016unsupervised,zhang2016colorful}. \textit{Contrastive} self-supervised methods such as instance discrimination~\cite{wu2018unsupervised}, CPC~\cite{henaff2019data,oord2018representation}, Deep InfoMax~\cite{hjelm2018learning}, Ye \textit{et al.}~\cite{ye2019unsupervised}, AMDIM~\cite{bachman2019learning}, CMC~\cite{tian2019contrastive}, MoCo~\cite{chen2020improved,he2020momentum}, PIRL~\cite{misra2020self}, and SimCLR~\cite{chen2020simple,chen2020big} were the first to achieve linear classification accuracy approaching that of end-to-end supervised training. Recently, these methods have been harnessed to achieve dramatic improvements in label efficiency for semi-supervised learning. Specifically, one can first pretrain in a task-agnostic, self-supervised fashion using all data, and then fine-tune on the labeled subset in a task-specific fashion with a standard supervised objective~\cite{chen2020simple,chen2020big,henaff2019data}. Chen \textit{et al.}~\cite{chen2020big} show that this approach benefits substantially from large (high-capacity) models for pretraining and fine-tuning, but after a large model is trained, it can be distilled to a much smaller model with little loss in accuracy.

Our Multi-Instance Contrastive Learning approach is related to previous works in video processing where multiple views naturally arising due to temporal variation~\cite{sermanet2018time,tschannen2020vivi}. These works have proposed to learn visual representations from video by maximizing agreement between representations of adjacent frames~\cite{tschannen2020vivi} or two views of the same action~\cite{sermanet2018time}. We generalize this idea to representation learning from image datasets, when sets of images containing the same desired class information are available and we show that benefits of MICLe can be combined with state-of-the-art self-supervised learning methods such as SimCLR.

\vspace{3pt}
\noindent {\bf Self-supervision for Medical Image Analysis.} Although self-supervised learning has only recently become viable on standard image classification datasets, it has already seen some application within the medical domain. While some works have attempted to design domain-specific pretext tasks~\cite{bai2019self,spitzer2018improving,zhuang2019self,zhu2020rubik}, other works concentrate on tailoring contrastive learning to medical data~\cite{he2020sample,li2021imbalance,liu2019align,zhou2020comparing}. Most closely related to our work, Sowrirajan \textit{et al.}~\cite{sowrirajan2020moco} explore the use of MoCo pretraining for classification of CheXpert dataset through linear evaluation. 

Several recent publications investigate semi-supervised learning for medical imaging tasks (\eg~\cite{cheplygina2019not,liu2020semi,wang2020focalmix,zhang2020contrastive}). These methods are complementary to ours, and we believe combining self-training and self-supervised pretraining is an interesting avenue for future research (\eg~\cite{chen2020big}).

\vspace{-4pt}
\section{Self-Supervised Pretraining}
\vspace{-4pt}
Our approach comprises the following steps. First, we perform self-supervised pretraining on unlabeled images using contrastive learning to learn visual representations. For contrastive learning we use a combination of unlabeled ImageNet dataset and task specific medical images. Then, if multiple images of each medical condition are available the Multi-Instance Contrastive Learning (MICLe) is used for additional self-supervised pretraining. Finally, we perform supervised fine-tuning on labeled medical images. Figure~\ref{fig:fig-1} shows the summary of our proposed method.

\subsection{A Simple Framework for Contrastive Learning}
To learn visual representations effectively with unlabeled images, we adopt SimCLR~\cite{chen2020simple,chen2020big}, a recently proposed approach based on contrastive learning. SimCLR learns representations by maximizing agreement~\cite{becker1992self} between differently augmented views of the same data example via a contrastive loss in a hidden representation of neural nets. 

Given a randomly sampled mini-batch of images, each image $\bm x_i$ is augmented twice using random crop, color distortion and Gaussian blur, creating two views of the same example $\bm x_{2k-1}$ and $\bm x_{2k}$. The two images are encoded via an encoder network $f(\cdot)$ (a ResNet~\cite{he2016deep}) to generate representations $\bm h_{2k-1}$ and $\bm h_{2k}$. The representations are then transformed again with a non-linear transformation network $g(\cdot)$ (a MLP projection head), yielding $\bm z_{2k-1}$ and $\bm z_{2k}$ that are used for the contrastive loss. 

With a mini-batch of encoded examples, the contrastive loss between a pair of positive example $i, j$ (augmented from the same image) is given as follows:
\begin{equation}
\label{eq:nt_xent}
    \ell^{\mathrm{NT}\text{-}\mathrm{Xent}}_{i,j} = -\log \frac{\exp(\mathrm{sim}(\bm z_i, \bm z_j)/\tau)}{\sum_{k=1}^{2N} \one{k \neq i}\exp(\mathrm{sim}(\bm z_i, \bm z_k)/\tau)}~,
\end{equation}
Where $\mathrm{sim}(\cdot,\cdot)$ is cosine similarity between two vectors, and $\tau$ is a temperature scalar.


\vspace{-2pt}
\subsection{Multi-Instance Contrastive Learning (MICLe)}
\vspace{-2pt}
\label{sec:micle}
In medical image analysis, it is common to utilize multiple images per patient to improve classification accuracy and robustness. Such images may be taken from different viewpoints or under different lighting conditions, providing complementary information for medical diagnosis. When multiple images of a medical condition are available as part of the training dataset, we propose to learn representations that are invariant not only to different augmentations of the same image, but also to different images of the same medical pathology. Accordingly, we can conduct a multi-instance contrastive learning (MICLe) stage where positive pairs are constructed by drawing two crops from the images of the same patient as demonstrated in \figref{fig:micle}. 

In MICLe, in contrast to standard SimCLR, to construct a mini-batch of $2N$ representation, we randomly sample a mini-batch of $N$ bags of instances and define the contrastive prediction task on positive pairs retrieved from the bag of images instead of augmented views of the same image. Each bag, $X=\{\bm x_1, \bm x_2, ..., \bm x_M\}$ contains images from a same patient (\ie~same pathology) captured from different views and we assume that $M$ could vary for different bags. When there is two or more instances in a bag ($M=\left|X\right|\geq2$), we construct positive pairs by drawing two crops from two randomly selected images in the bag. In this case, the objective still takes the form of Eq.~\eqref{eq:nt_xent}, but images contributing to each positive pair are distinct. Algorithm~\ref{alg:micle} summarizes the proposed method.

Leveraging multiple images of the same condition using the contrastive loss helps the model learn representations that are more robust to the change of viewpoint, lighting conditions, and other confounding factors. We find that multi-instance contrastive learning significantly improves the accuracy and helps us achieve the state-of-the-art result on the dermatology condition classification task.

\begin{algorithm}[h]
\SetAlgoLined
\small{
\KwInput{batch size $N$, constant $\tau$, $g(\cdot)$, $f(\cdot)$, $\mathcal{T}$}
    \While{stopping criteria not met}{
        Sample mini-batch of $\{X\}_{k=1}^N$
        \For{$k\gets1$ \textbf{to} $k=N$}{
            Draw augmentation functions $t$ and $t'$ $\sim$ $\mathcal{T}$\;
            \eIf{$\left|X_k\right|\geq2$}{
                Randomly select $\bm x_k$ and $\bm x'_k$ $\in X_k$\;
            }{
                $\bm x_k = \bm x'_k\gets$ the only element of $X_k$\;
            }
            $\tilde{\bm x}_{2k-1} = t(\bm x_k)$; $\tilde{\bm x}_{2k} = t'(\bm x'_k)$\;
            $\bm z_{2k-1} = g(f(\tilde{\bm x}_{2k-1}))$; $\bm z_{2k} = g(f(\tilde{\bm x}_{2k}))$\;
        }
        \For{$i\in\{1, \ldots, 2N\}$ and $j\in\{1, \dots, 2N\}$}{
            $s_{i,j} = \bm z_i^\top \bm z_j / (\lVert\bm z_i\rVert \lVert\bm z_j\rVert)$\;
        }
        $\ell(i, j) \gets \ell^{\mathrm{NT}\text{-}\mathrm{Xent}}_{i,j}$ in Eq.~\eqref{eq:nt_xent}\;
        $\mathcal{L} = \frac{1}{2N} \sum_{k=1}^N \left[ \ell(2k\!-\!1, 2k) + \ell(2k, 2k\!-\!1)\right]$\;
    }
    \textbf{return} Trained encoder network $f(\cdot)$
}
\caption{\small{Multi-Instance Contrastive Learning.}}
\label{alg:micle}
\end{algorithm}

\vspace{-4pt}
\section{Experiment Setup} 
\vspace{-2pt}

\newcommand{\datasetdermtrain}{$\mathcal{D}_{\text{Derm}}^{\text{Train}}$}
\newcommand{\datasetdermvalidtaion}{$\mathcal{D}_{\text{Derm}}^{\text{Validation}}$}
\newcommand{\datasetdermtest}{$\mathcal{D}_{\text{Derm}}^{\text{Test}}$}
\newcommand{\datasetderm}{$\mathcal{D}_{\text{Derm}}$}
\newcommand{\datasetdermexternal}{$\mathcal{D}_{\text{Derm}}^{\text{External}}$}
\newcommand{\datasetdermunlabeled}{$\mathcal{D}_{\text{Derm}}^{\text{Unlabeled}}$}

\newcommand{\datasetchexunlabeled}{$\mathcal{D}_{\text{CheXpert}}^{\text{Unlabeled}}$}
\newcommand{\datasetnih}{$\mathcal{D}_{\text{NIH}}^{\text{}}$}

\subsection{Tasks and datasets}
We consider two popular medical imaging tasks. The first task is in the dermatology domain and involves identifying skin conditions from digital camera images. The second task involves multi-label classification of chest X-rays among five pathologies. We chose these tasks as they embody many common characteristics of medical imaging tasks like imbalanced data and pathologies of interest restricted to small local patches. At the same time, they are also quite diverse in terms of the type of images, label space and task setup. For example, dermatology images are visually similar to natural images whereas the chest X-rays are gray-scale and have standardized views. This, in turn, helps us probe the generality of our proposed methods.

\noindent \textbf{Dermatology. }
For the dermatology task, we follow the experiment setup and dataset of~\cite{liu2020deep}.  The dataset was collected and  de-identified by a US based tele-dermatology service with images of skin conditions taken using consumer grade digital cameras. The images are heterogeneous in nature and exhibit significant variations in terms of the pose, lighting, blur, and body parts. The background also contains various noise artifacts like clothing and walls which adds to the challenge. The ground truth labels were aggregated from a panel of several US-board certified dermatologists who provided differential diagnosis of skin conditions in each case. 

In all, the dataset has cases from a total of 12,306 unique patients. Each case includes between one to six images. This further split into development and test sets ensuring no patient overlap between the two. Then, cases with the occurrence of multiple skin conditions or poor quality images were filtered out. The final~\datasetdermtrain,~\datasetdermvalidtaion, and~\datasetdermtest  include a total of 15,340 cases, 1,190 cases, and 4,146 cases, respectively. There are 419 unique condition labels in the dataset. For the purpose of model development, we identified and use the most common 26 skin conditions and group the rest in an additional `Other' class leading to a final label space of 27 classes for the model. We refer to this as \datasetderm in the subsequent sections. We also use an additional de-identified \datasetdermexternal set to evaluate the generalization performance of our proposed method under distribution shift. Unlike \datasetderm, this dataset is primarily focused on skin cancers and the ground truth labels are obtained from biopsies. The distribution shift in the labels make this a particular challenging data to evaluate the zero-shot (i.e. without any additional fine-tuning) transfer performance of models. 

For SimCLR pretraining, we combine the images from \datasetdermtrain~ and additional unlabeled images from the same source leading to a total of 454,295 images for self-supervised pretraining. We refer to this as the \datasetdermunlabeled. For MICLe pretraining, we only use images coming from the 15,340 cases of the \datasetdermtrain. Additional details are provided in the Appendix~\ref{app:datasets-derm}.

\vspace{2pt}
\noindent \textbf{Chest X-rays. }
CheXpert~\cite{irvin2019chexpert} is a large open source dataset of de-identified chest radiograph (X-ray) images. The dataset consists of a set of 224,316 chest radiographs coming from 65,240 unique patients. The ground truth labels were automatically extracted from radiology reports and correspond to a label space of 14 radiological observations. The validation set consists of 234 manually annotated chest X-rays. Given the small size of the validation dataset and following~\cite{neyshabur2020being,raghu2019transfusion} suggestion, for the downstream task evaluations we randomly re-split the training set into 67,429 training images, 22,240 validation images, and 33,745 test images. We train the model to predict the five pathologies used by Irvin and Rajpurkar \textit{et al.}~\cite{irvin2019chexpert} in a multi-label classification task setting. For SimCLR pretraining for the chest X-ray domain, we only consider images coming from the train set of the CheXpert dataset discarding the labels. We refer to this as the \datasetchexunlabeled. In addition, we also use the NIH chest X-ray dataset, \datasetnih, to evaluate the zero-shot transfer performance which consist of 112,120 de-identified X-rays from 30,805 unique patients. Additional details on the dataset can be found here~\cite{wang2017chestx} and also are provided in the Appendix~\ref{app:datasets-chex}.

\vspace{-3pt}
\subsection{Pretraining protocol}
\vspace{-3pt}
To assess the effectiveness of self-supervised pretraining using big neural nets, as suggested in~\cite{chen2020simple}, we investigate  ResNet-50 (1$\!\times$), ResNet-50 (4$\!\times$), and ResNet-152 (2$\!\times$) architectures as our base encoder networks. Following SimCLR~\cite{chen2020simple}, two fully connected layers are used to map the output of ResNets to a 128-dimensional embedding, which is used for contrastive learning. We also use LARS optimizer~\cite{you2017large} to stabilize training during pretraining. We perform SimCLR pretraining on \datasetdermunlabeled~and \datasetchexunlabeled, both with and without initialization from ImageNet self-supervised pretrained weights. We indicate pretraining initialized using self-supervised ImageNet weights, as ImageNet~$\!\rightarrow\!$~Derm, and ImageNet~$\!\rightarrow\!$~CheXpert in the following sections.

Unless otherwise specified, for the dermatology pretraining task, due to similarity of dermatology images to natural images, we use the same data augmentation used to generate positive pairs in SimCLR. This includes random color augmentation (strength=1.0), crops with resize, Gaussian blur, and random flips. We find that the batch size of 512 and learning rate of 0.3 works well in this setting. Using this protocol, all of models were pretrained up to 150,000 steps using \datasetdermunlabeled. 

For the CheXpert dataset, we pretrain with learning rate in $\{$0.5, 1.0, 1.5$\}$, temperature in $\{$0.1, 0.5, 1.0$\}$, and batch size in $\{$512, 1024$\}$, and we select the model with best performance on the down-stream validation set. We also tested a range of possible augmentations and observe that the augmentations that lead to the best performance on the validation set for this task are random cropping, random color jittering ($\text{strength}=0.5$),  rotation (upto 45 degrees) and horizontal flipping. Unlike the original set of proposed augmentation in SimCLR, we do not use the Gaussian blur, because we think it can make it impossible to distinguish local texture variations and other areas of interest thereby changing the underlying disease interpretation the X-ray image. We leave comprehensive investigation of the optimal augmentations to future work. Our best model on CheXpert was pretrained with batch size 1024, and learning rate of 0.5 and we pretrain the models up to 100,000 steps.

We perform \multisim~pretraining only on the dermatology dataset as we did not have enough cases with the presence of multiple views in the CheXpert dataset to allow comprehensive training and evaluation of this approach. For \multisim~pretraining we initialize our model using SimCLR pretrained weights, and then incorporate the multi-instance procedure as explained in Section~\ref{sec:micle} to further learn a more comprehensive representation using multi-instance data of \datasetdermtrain. Due to memory limits caused by stacking up to 6 images per patient case, we train with a smaller batch size of 128 and learning rate of 0.1 for 100,000 steps to stabilize the training. Decreasing the learning rate for smaller batch size has been suggested in~\cite{chen2020simple}. The rest of the settings, including optimizer, weight decay, and warmup step are the same as our previous pretraining protocol.

In all of our pretraining experiments, images are resized to 224 $\times$ 224. We use 16 to 64 Cloud TPU cores depending on the batch size for pretraining. With 64 TPU cores, it takes $\sim$12 hours to pretrain a ResNet-50 (1$\!\times$) with batch size 512 and for 100 epochs. Additional details about the selection of batch size and learning rate, and augmentations are provided in the Appendix~\ref{app:technical}.


\begin{table*}[!t]
\centering
\vspace{-5pt}
\caption{\small Performance of dermatology skin condition and Chest X-ray classification model measured by top-1 accuracy (\%) and area under the curve (AUC) across different architectures. Each model is fine-tuned using transfer learning from pretrained model on ImageNet, only unlabeled medical data, or pretrained using medical data initialized from ImageNet pretrained model (e.g. ImageNet~$\!\rightarrow\!$~Derm). Bigger models yield better performance. pretraining on ImageNet is complementary to pretraining on unlabeled medical images.}\label{tab: 1 }
\vspace{-7pt}
\small
\begin{tabular}{l|ccc|ccc}\toprule
 &\multicolumn{3}{c|}{Dermatology Classification} &\multicolumn{2}{c}{Chest X-ray Classifcation}                             \\\midrule
Architecture &~~Pretraining Dataset~~ &~~Top-1 Accuracy(\%)~~  &~~AUC~~  &~~Pretraining Dataset~~ &~~~ Mean AUC~~~           \\ \midrule
\multirow{3}{*}{ResNet-50 (1$\!\times$)} 
&ImageNet                       &62.58 $\pm$ 0.84   &0.9480 $\pm$ 0.0014 &ImageNet                       &0.7630 $\pm$ 0.0013 \\
&Derm                           &63.66 $\pm$ 0.24   &0.9490 $\pm$ 0.0011 &CheXpert                       &0.7647 $\pm$ 0.0007 \\ 
&ImageNet$\rightarrow$Derm      &63.44 $\pm$ 0.13   &0.9511 $\pm$ 0.0037 &ImageNet$\rightarrow$CheXpert  &0.7670 $\pm$ 0.0007 \\ \midrule
\multirow{3}{*}{ResNet-50 (4$\!\times$)} 
&ImageNet                       &64.62 $\pm$ 0.76   &0.9545 $\pm$ 0.0007 &ImageNet                       &0.7681 $\pm$ 0.0008 \\
&Derm                           &66.93 $\pm$ 0.92   &0.9576 $\pm$ 0.0015 &CheXpert                       &0.7668 $\pm$ 0.0011 \\
&ImageNet$\rightarrow$Derm      &67.63 $\pm$ 0.32   &0.9592 $\pm$ 0.0004 &ImageNet$\rightarrow$CheXpert  &0.7687 $\pm$ 0.0016 \\ \midrule
\multirow{3}{*}{ResNet-152 (2$\!\times$)} 
&ImageNet                       &66.38 $\pm$ 0.03   &0.9573 $\pm$ 0.0023 &ImageNet                      &0.7671 $\pm$ 0.0008 \\
&Derm                           &66.43 $\pm$ 0.62   &0.9558 $\pm$ 0.0007 &CheXpert                      &0.7683 $\pm$ 0.0009 \\
&ImageNet$\rightarrow$Derm      &68.30 $\pm$ 0.19   &0.9620 $\pm$ 0.0007 &ImageNet$\rightarrow$CheXpert &0.7689 $\pm$ 0.0010 \\
\bottomrule
\end{tabular}
\vspace{-12pt}
\end{table*}

\vspace{-3pt}
\subsection{Fine-tuning protocol}
\vspace{-3pt}
We train the model end-to-end during fine-tuning using the weights of the pretrained network as initialization for the downstream supervised task dataset following the approach described by Chen \textit{et al.}~\cite{chen2020simple,chen2020big} for all our experiments. We trained for 30,000 steps with a batch size of 256 using SGD with a momentum parameter of 0.9. For data augmentation during fine-tuning, we performed random color augmentation, crops with resize, blurring, rotation, and flips for the images in both tasks. We observe that this set of augmentations is critical for achieving the best performance during fine-tuning. We resize the Derm dataset images to 448~$\times$~448 pixels and CheXpert images to 224~$\times$~224 during this fine-tuning stage.

For every combination of pretraining strategy and downstream fine-tuning task, we perform an extensive hyper-parameter search. We selected the learning rate and weight decay after a grid search of seven logarithmically spaced learning rates between 10$^{-3.5}$ and 10$^{-0.5}$ and three logarithmically spaced values of weight decay between 10$^{-5}$ and 10$^{-3}$, as well as no weight decay. For training from the supervised pretraining baseline we follow the same protocol and observe that for all fine-tuning setups, 30,000 steps is sufficient to achieve optimal performance. For supervised baselines we compare against the identical publicly available ResNet models\footnote{https://github.com/google-research/simclr} pretrained on ImageNet with standard cross-entropy loss. These models are trained with the same data augmentation as self-supervised models (crops, strong color augmentation, and blur).

\vspace{-2pt}
\subsection{Evaluation methodology} 
\vspace{-2pt}
After identifying the best hyperparameters for fine-tuning a given dataset, we proceed to select the model based on validation set performance and evaluate the chosen model multiple times (10 times for chest X-ray task and 5 times for the dermatology task) on the test set to report task performance. Our primary metrics for the dermatology task are top-1 accuracy and Area Under the Curve (AUC) following~\cite{liu2020deep}. For the chest X-ray task, given the multi-label setup, we report mean AUC averaged between the predictions for the five target pathologies following~\cite{irvin2019chexpert}. We also use the non-parametric bootstrap to estimate the variability around the model performance and investigating any statistically significant improvement. Additional details are provided in Appendix~\ref{app:technical-stat}.

\vspace{-3pt}
\section{Experiments \& Results} 
\vspace{-3pt}
In this section we investigate whether self-supervised pretraining with contrastive learning translates to a better performance in models fine-tuned end-to-end across the selected medical image classification tasks. To this end, first, we explore the choice of the pretraining dataset for medical imaging tasks. Then, we evaluate the benefits of our proposed multi-instance contrastive learning (\multisim) for dermatology condition classification task, and compare and contrast the proposed method against the baselines and state of the art methods for supervised pretraining. Finally, we explore label efficiency and transferability (under distribution shift) of self-supervised trained models in the medical image classification setting.

\vspace{-1pt}
\subsection{Dataset for pretraining}
\vspace{-1pt}
One important aspect of transfer learning via self-supervised pretraining is the choice of a proper unlabeled dataset. For this study, we use architectures of varying capacities (i.e ResNet-50 (1$\!\times$), ResNet-50 (4$\!\times$) and ResNet-152 (2$\!\times$) as our base network, and carefully investigate three possible scenario for self-supervised pretraining in the medical context: (1) using ImageNet dataset only , (2) using the task specific unlabeled medical dataset (i.e. Derm and CheXpert), and (3) initializing the pretraining from ImageNet self-supervised model but using task specific unlabeled dataset for pretraining, here indicated as ImageNet~$\!\rightarrow\!$~CheXpert and ImageNet~$\!\rightarrow\!$~CheXpert. Table~\ref{tab: 1 } shows the performance of dermatology skin condition and chest X-ray classification model measured by top-1 accuracy (\%) and area under the curve (AUC) across different architectures and pretraining scenarios. Our results suggest that, best performance are achieved when both ImageNet and task specific unlabeled data are used. Combining ImageNet and Derm unlabeled data for pretraining, translates to $(1.92\pm0.16)\%$ increase in top-1 accuracy for dermatology classification over only using ImageNet dataset for self-supervised transfer learning.  This results suggests that pretraining on ImageNet is likely complementary to pretraining on unlabeled medical images. Moreover, we observe that larger models are able to benefit much more from self-supervised pretraining underscoring the importance of model capacity in this setting.

As shown in Table~\ref{tab: 1 }, on CheXpert, we once again observe that self-supervised pretraining with both ImageNet and in-domain CheXpert data is beneficial, outperforming self-supervised pretraining on ImageNet or CheXpert alone.

\vspace{-1pt}
\subsection{Performance of MICLe}
\vspace{-1pt}
Next, we evaluate whether utilizing multi-instance contrastive learning (MICLe) and leveraging the potential availability of multiple images per patient for a given pathology, is beneficial for self-supervised pretraining. Table~\ref{tab: 2} compares the performance of dermatology condition classification models fine-tuned on representations learned with and without MICLe pretraining. We observe that MICLe consistently improves the performance of dermatology classification over the original SimCLR method under different pretraining dataset and base network architecture choices.  Using MICLe for pretraining, translates to $(1.18\pm0.09)\%$ increase in  top-1  accuracy  for  dermatology classification over using only original SimCLR. 

\begin{table}[t]\centering
\caption{\small Evaluation of multi instance contrastive learning (\multisim) on Dermatology condition classification. Our results suggest that MICLe consistently improves the  accuracy  of  skin condition  classification  over SimCLR on different datasets and architectures.
}\label{tab: 2}
\vspace{-7pt}
\small
\begin{tabular}{@{}c|cc|c}\toprule
Model          &Dataset                   &MICLe                  &Top-1 Accuracy          \\\midrule
               &Derm                      &No                     &66.93$\pm$0.92          \\
ResNet-50      &Derm                      &Yes                    &67.55$\pm$0.52           \\
(4$\!\times$)  &ImageNet$\rightarrow$Derm &No                     &67.63$\pm$0.32          \\
               &ImageNet$\rightarrow$Derm &\textbf{Yes}           &\textbf{68.81}$\pm$0.41 \\ \midrule
               &Derm                      &No                     &66.43$\pm$0.62          \\
ResNet-152     &Derm                      &Yes                    &67.16$\pm$0.35          \\
(2$\!\times$)  &ImageNet$\rightarrow$Derm &No                     &68.30$\pm$0.19          \\
               &ImageNet$\rightarrow$Derm &\textbf{Yes}           &\textbf{68.43}$\pm$0.32 \\
\bottomrule
\end{tabular}
\end{table}

\vspace{-2pt}
\subsection{Comparison with supervised transfer learning}
\vspace{-2pt}
We further improves the performance by providing more negative examples with training longer for 1000 epochs and a larger batch size of 1024. We achieve the best-performing top-1 accuracy of $(70.02\pm0.22)$\% using the ResNet-152 (2$\!\times$) architecture and MICLe pretraining by incorporating both ImageNet and Derm dataset in dermatology condition classification. Tables~\ref{tab: 3} and~\ref{tab: 4} show the comparison of transfer learning performance of SimCLR and MICLe models with supervised baselines for the dermatology and the chest X-ray classification. This result shows that after fine-tuning, our self-supervised model significantly outperforms the supervised baseline when ImageNet pretraining is used ($p<0.05$). We specifically observe an improvement of over 6.7\% in top-1 accuracy in the dermatology task when using MICLe. On the chest X-ray task, the improvement is 1.1\% in mean AUC without using MICLe. 

Though using ImageNet pretrained models is still the norm, recent advances have been made by supervised pretraining on large scale (often noisy) natural datasets~\cite{kolesnikov2019big,mahajan2018exploring} improving transfer performance on downstream tasks. We therefore also evaluate a supervised baseline from Kolesnikov \textit{et al.}~\cite{kolesnikov2019big}, a ResNet-101 (3$\!\times$) pretrained on ImageNet21-k called Big Transfer (BiT). This model contains additional architectural tweaks included to boost transfer performance, and was trained on a significantly larger dataset (14M images labelled with one or more of 21k classes, \vs{} the 1M images in ImageNet) which provides us with a strong supervised baseline\footnote{This model is also available publicly at \url{https://github.com/google-research/big_transfer}}. ResNet-101 (3$\!\times$) has 382M trainable parameters, thus comparable to ResNet-152 (2$\!\times$) with 233M trainable parameters. We observe that the MICLe model is better than this BiT model for the dermatology classification task improving by 1.6\% in top-1 accuracy. For the chest X-ray task, self supervised model is better by about 0.1\% mean AUC. We surmise that with additional in-domain unlabeled data (we only use the CheXpert dataset for pretraining), self-supervised pretraining can surpass the BiT baseline by a larger margin. At the same time, these two approaches are complementary but we leave further explorations in this direction to future work.

\begin{table}[!t]\centering
\caption{\small Comparison of best self-supervised models \vs{} supervised pretraining baselines on dermatology classification.}\label{tab: 3}
\vspace{-7pt}
\small
\begin{tabular}{@{}l@{\hspace{.2cm}}c@{}c@{\hspace{.2cm}}c@{}}\toprule
Architecture &Method &Pretraining Dataset &Top-1 Accuracy \\\midrule
ResNet-152 (2$\!\times$) &Supervised    &ImageNet         &63.36 $\pm$ 0.12 \\
ResNet-101 (3$\!\times$) &BiT~\cite{kolesnikov2019big} &ImageNet-21k &68.45 $\pm$ 0.29 \\
\midrule
ResNet-152 (2$\!\times$) &SimCLR        &ImageNet         &66.38 $\pm$ 0.03  \\
ResNet-152 (2$\!\times$) &SimCLR        &ImageNet~$\!\rightarrow\!$~Derm  &69.43 $\pm$ 0.43 \\
ResNet-152 (2$\!\times$) &\multisim     &ImageNet~$\!\rightarrow\!$~Derm  &{\bf 70.02} $\pm$ 0.22 \\
\bottomrule
\end{tabular}
\vspace{-4pt}
\end{table}

\begin{table}[!t]\centering
\caption{\small Comparison of best self-supervised models \vs{} supervised pretraining baselines on chest X-ray classification.}\label{tab: 4}
\vspace{-7pt}
\small
\begin{tabular}{@{}l@{\hspace{.05cm}}c@{\hspace{.1cm}}c@{\hspace{.2cm}}c@{}}\toprule
Architecture &Method &Pretraining Dataset &Mean AUC \\\midrule
ResNet-152 (2$\!\times$) &Supervised   &ImageNet                     &0.7625 $\pm$ 0.001 \\
ResNet-101 (3$\!\times$) &BiT~\cite{kolesnikov2019big} &ImageNet-21k &0.7720 $\pm$ 0.002  \\
\midrule
ResNet-152 (2$\!\times$) &SimCLR       &ImageNet                     &0.7671 $\pm$ 0.008 \\
ResNet-152 (2$\!\times$) &SimCLR       &CheXpert                     &0.7702 $\pm$ 0.001 \\
ResNet-152 (2$\!\times$) &SimCLR       &ImageNet~$\!\rightarrow\!$~CheXpert            &{\bf 0.7729} $\pm$ 0.001 \\
\bottomrule
\end{tabular}
\end{table}

\vspace{-1pt}
\subsection{Self-supervised models generalize better}
\vspace{-1pt}
We conduct further experiments to evaluate the robustness of self-supervised pretrained models to distribution shifts. For this purpose, we use the model post pretraining and end-to-end fine-tuning  (i.e. CheXpert and Derm) to make predictions on an additional shifted dataset without any further fine-tuning (zero-shot transfer learning). We use the \datasetdermexternal and \datasetnih~as our target shifted datasets. Our results generally suggest that self-supervised pretrained models can generalize better to distribution shifts.

For the chest X-ray task, we note that self-supervised pretraining with either ImageNet or CheXpert data improves generalisation, but stacking them both yields further gains. We also note that when only using ImageNet for self supervised pretraining, the model performs worse in this setting compared to using in-domain data for pretraining. 

Further we find that the performance improvement in the distribution-shifted dataset due to self-supervised pretraining (both using ImageNet and CheXpert data) is more pronounced than the original improvement on the CheXpert dataset. This is a very valuable finding, as generalisation under distribution shift is of paramount importance to clinical applications. On the dermatology task, we observe similar trends suggesting the robustness of the self-supervised representations is consistent across tasks.

\begin{figure}[t]
     \centering
         \includegraphics[height=0.245\textheight]{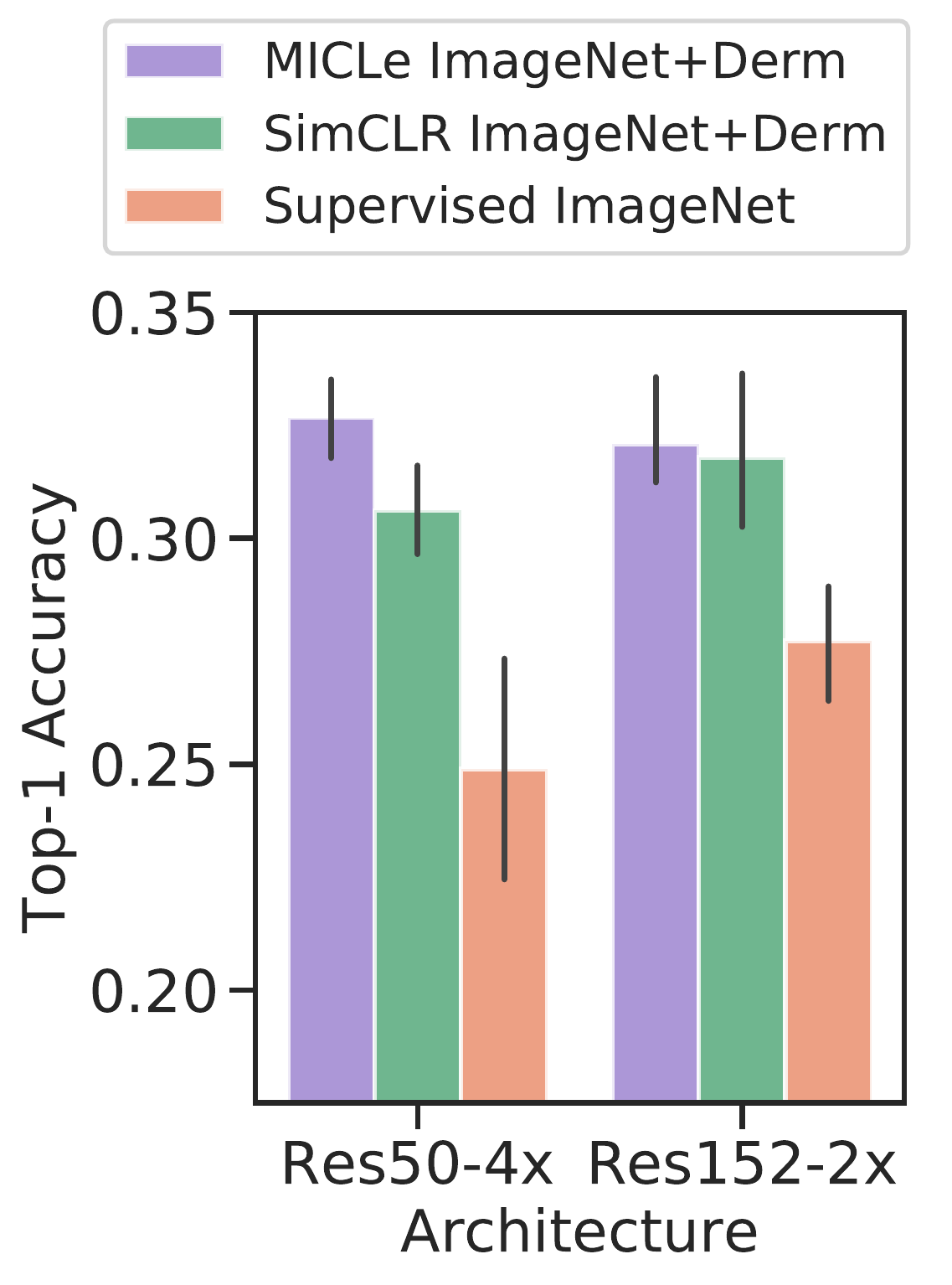}
         \centering
         \includegraphics[height=0.245\textheight]{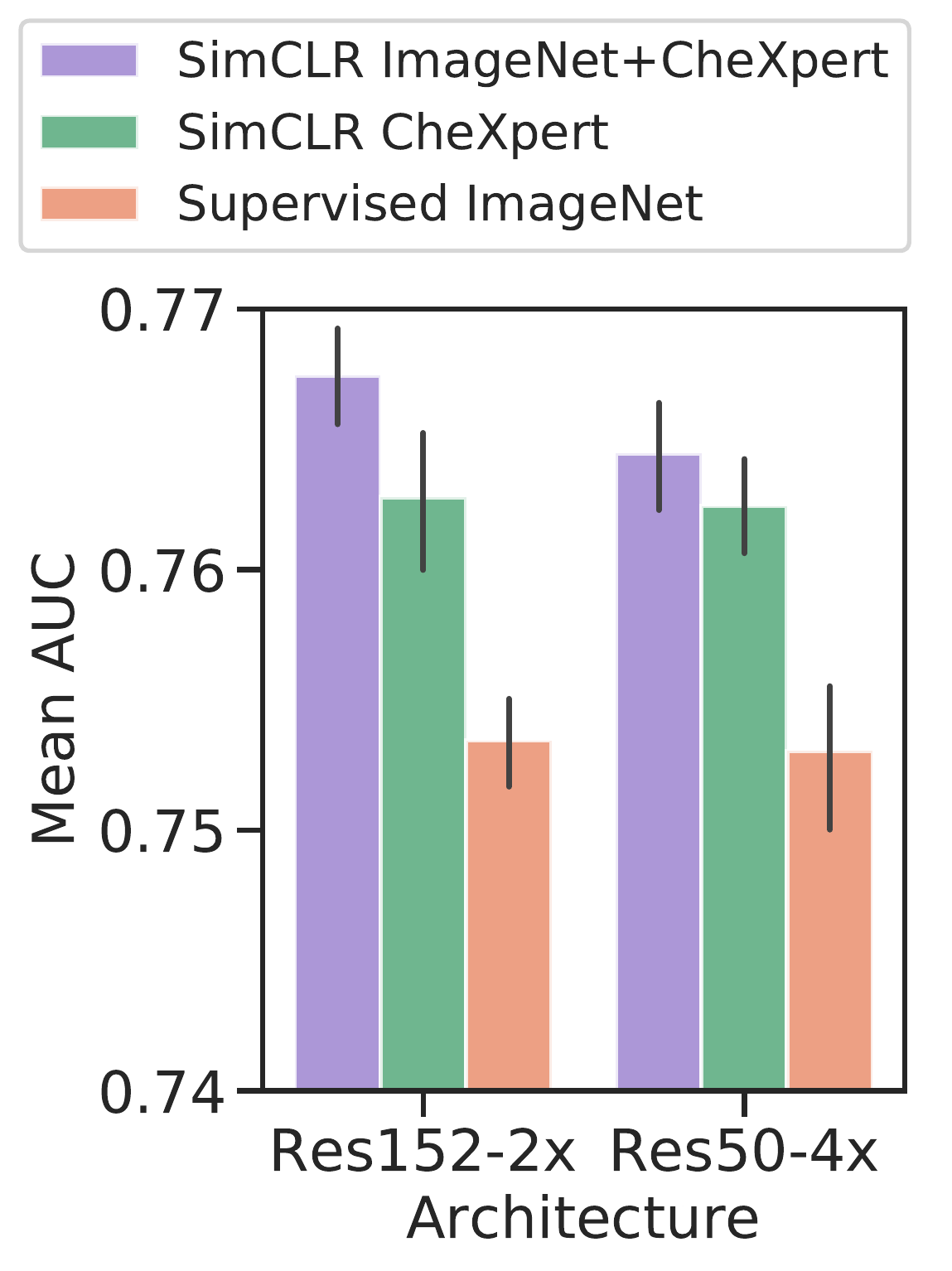}
     \hfill
        \vspace{-.55cm}
        \caption{\small Evaluation of models on distribution-shifted datasets (left: \datasetdermunlabeled$\to$\datasetdermexternal; right: \datasetchexunlabeled$\to$\datasetnih) shows that self-supervised training \textbf{using both ImageNet and the target domain} significantly improves robustness to distribution shift.}
        \label{fig:fig-4}
        \vspace{-0cm}
\end{figure}

\vspace{-1pt}
\subsection{Self-supervised models are more label-efficient}
\vspace{-1pt}
To investigate label-efficiency of the selected self-supervised models, following the previously explained fine-tuning protocol, we fine-tune our models on different fractions of labeled training data. We also conduct baseline fine-tuning experiments with supervised ImageNet pretrained models.  We use the label fractions ranging from 10\% to 90\% for both Derm and CheXpert training datasets. Fine-tuning experiments on label fractions are repeated multiple times using the best parameters and averaged. Figure~\ref{fig:fig-4} shows how the performance varies using the different available label fractions for the dermatology task. First, we observe that pretraining using self-supervised models can significantly help with label efficiency for medical image classification, and in all of the fractions, self-supervised models outperform the supervised baseline. Moreover, these results suggest that MICLe yields proportionally larger gains when fine-tuning with fewer labeled examples. In fact, MICLe is able to match baselines using only 20\% of the training data for ResNet-50 (4$\!\times$) and 30\% of the training data for ResNet-152 (2$\!\times$). Results on the CheXpert dataset are included in Appendix~\ref{app:technical-chex}, where we observe a similar trend.

\begin{figure}[t]
     \centering
         \centering
         \includegraphics[width=.34\textwidth]{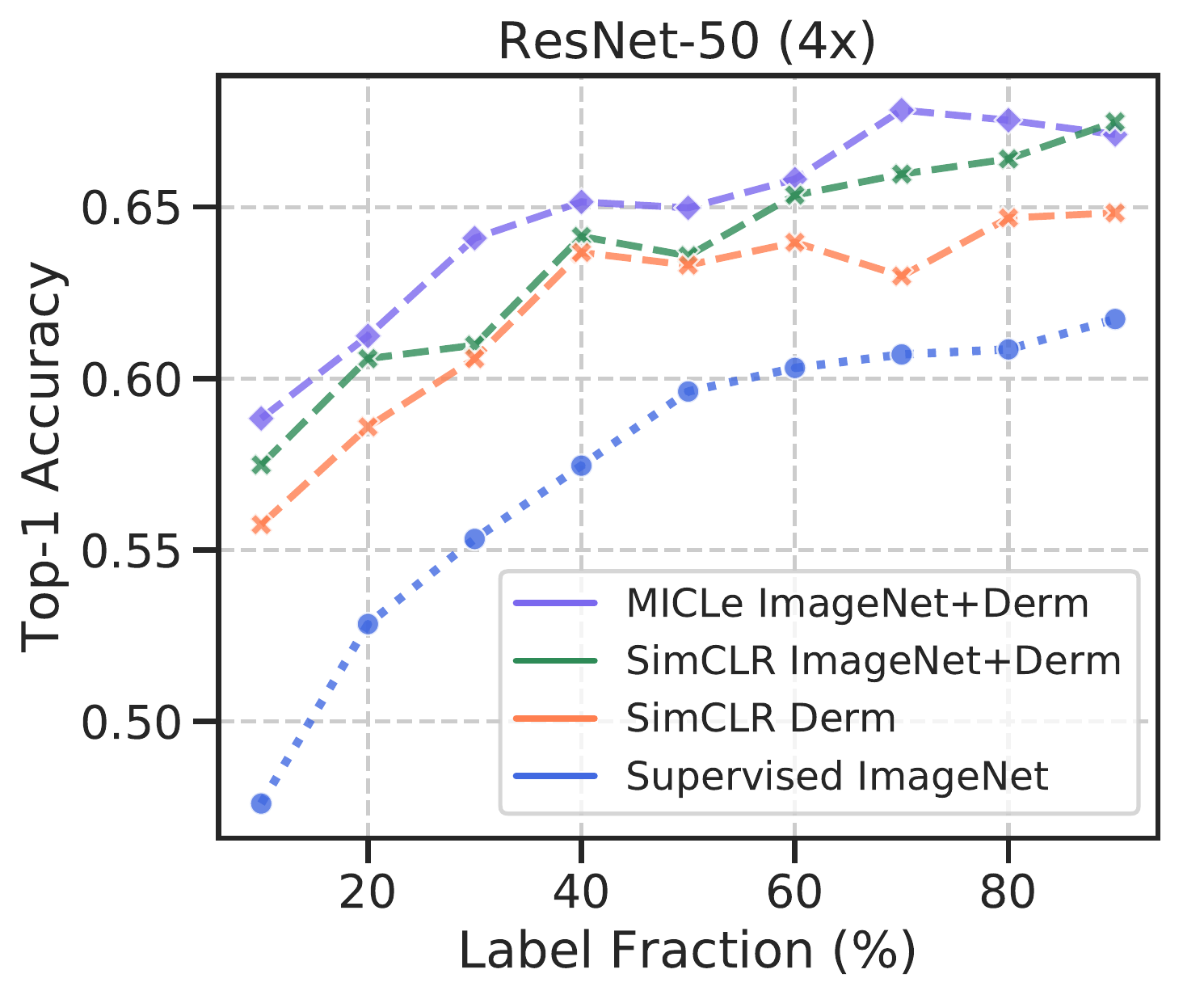}
     \hfill
         \centering
         \includegraphics[width=.34\textwidth]{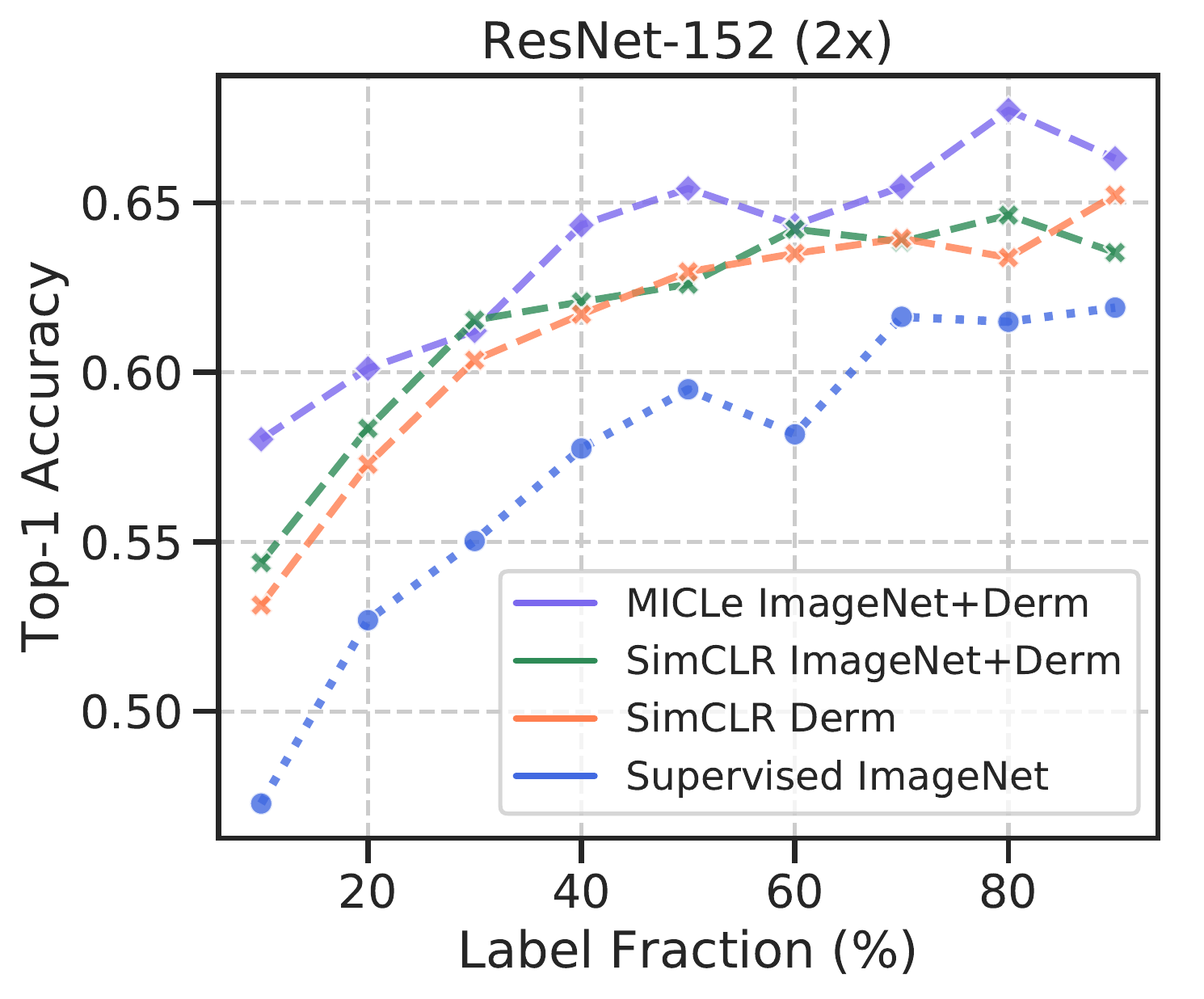}
        \vspace{-.4cm}
        \caption{\small Top-1 accuracy for dermatology condition classification for MICLe, SimCLR, and supervised models under different unlabeled pretraining dataset and varied sizes of label fractions. }
        \label{fig:fig-5}
        \vspace{-0cm}
\end{figure}

\vspace{-5pt}
\section{Conclusion}
\vspace{-5pt}
Supervised pretraining on natural image datasets such as ImageNet is commonly used to improve medical image classification. This paper investigates an alternative
strategy based on self-supervised pretraining on unlabeled natural and medical images and finds that self-supervised pretraining significantly outperforms supervised pretraining. The paper proposes the use of multiple images per medical case to enhance data augmentation for self-supervised learning, which boosts the performance of image classifiers even further. Self-supervised pretraining is much more scalable than supervised pretraining since class label annotation is not required. A natural next step for this line of research is to investigate the limit of self-supervised pretraining by considering massive unlabeled medical image datasets. Another research direction concerns the transfer of self-supervised learning from one imaging modality and task to another. We hope this paper will help popularize the use of self-supervised approaches in medical image analysis yielding improvements in label efficiency across the medical field.

\ificcvfinal{
    \section*{Acknowledgement}
    We would like to thank Yuan Liu for valuable feedback on the manuscript. We are also grateful to Jim Winkens, Megan Wilson, Umesh Telang, Patricia Macwilliams, Greg Corrado, Dale Webster, and our collaborators at DermPath AI for their support of this work. 
}\else{}\fi


{\small
\bibliographystyle{ieee_fullname}
\bibliography{main}
\balance
}

\appendix
\clearpage
\onecolumn

\appendix
\renewcommand{\thefigure}{\thesection.\arabic{figure}}
\renewcommand{\thetable}{\thesection.\arabic{table}} 
\renewcommand{\theequation}{\thesection.\arabic{equation}} 
\setcounter{figure}{0}
\setcounter{table}{0}
\setcounter{equation}{0}

\section{Datasets}
\label{app:datasets}
\subsection{Dermatology}
\label{app:datasets-derm}

\noindent \textbf{Dermatology dataset details.}
 As in actual clinical settings, the distribution of different skin conditions is heavily skewed in the Derm dataset, ranging from some skin conditions making up more than 10\% of the training data like acne, eczema, and psoriasis, to those making up less than 1\% like lentigo, melanoma, and stasis dermatitis~\cite{liu2020deep}. To ensure that there was sufficient data to develop and evaluate the Dermatology skin condition classifier, we filtered the 419 conditions to the top 26 with the highest prevalence based on the training set. Specifically, this ensured that for each of these conditions, there were at least 100 cases in the training dataset. The remaining conditions were aggregated into an ``Other" category (which comprised 21\% of the cases in test dataset).  The 26 target skin conditions are as follow: Acne, Actinic keratosis, Allergic contact dermatitis, Alopecia areata, Androgenetic alopecia, Basal cell carcinoma, Cyst, Eczema, Folliculitis, Hidradenitis, Lentigo, Melanocytic nevus, Melanoma, Post inflammatory hyperpigmentation, Psoriasis, Squamous cell carcinoma/squamous cell carcinoma insitu (SCC$/$SCCIS), Seborrheic keratosis, Scar condition, Seborrheic dermatitis, Skin tag, Stasis dermatitis, Tinea, Tinea versicolor, Urticaria, Verruca vulgaris, Vitiligo.

\begin{figure}[ht]
     \centering
     \includegraphics[width=0.6\textwidth]{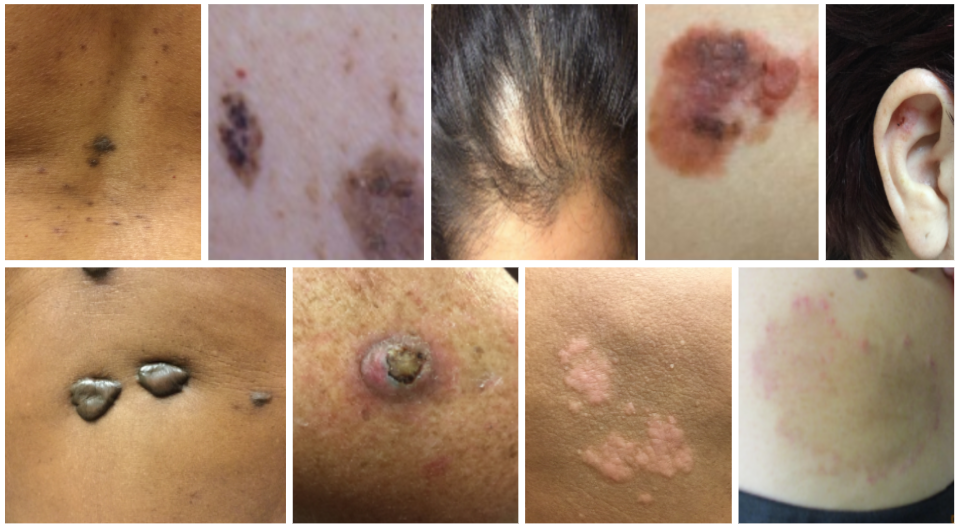}
    \caption{\small{Examples images from Derm dataset. Derm dataset includes 26 classes, ranging from skin conditions with greater than 10\% prevalence like acne, eczema, and psoriasis, to those with sub-1\% prevalence like lentigo, melanoma, and stasis dermatitis.}}
    \label{fig-sm:derm-sample}
\end{figure}

\begin{figure}[ht]
     \centering
     \includegraphics[width=0.6\textwidth]{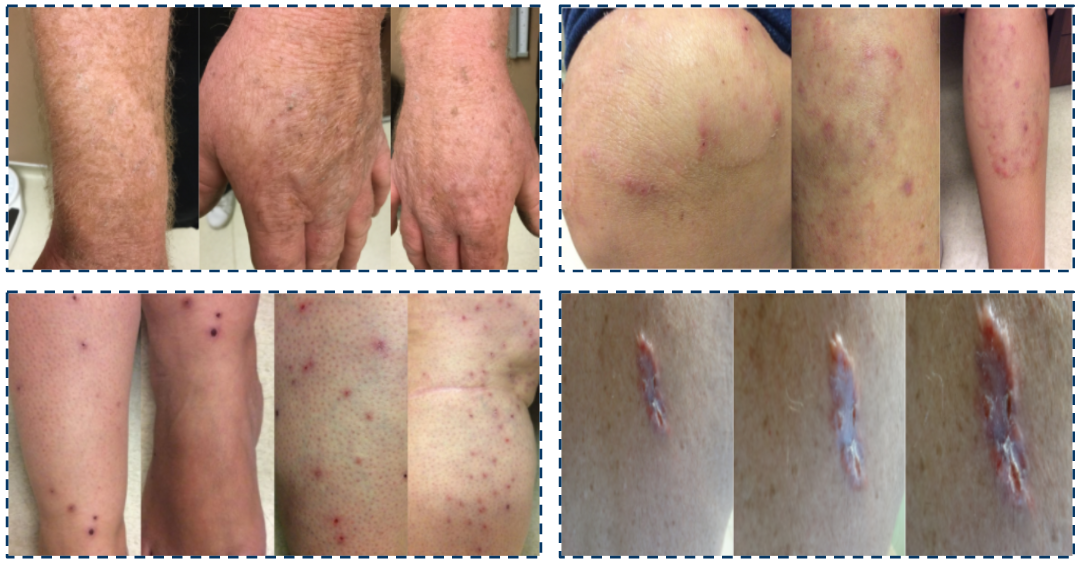}
    \caption{\small{Examples of images belong to the same patient which are taken from different viewpoints and/or from different body-parts under different lighting conditions. Each category, marked with a dashed line, belongs to a single patient and represents a single medical condition. In MICLe, when multiple images of a medical condition from the same patient are available, we use two randomly selected images from the patient to directly create a positive pair of examples and later adopt the augmentation. When a single image of a medical condition is available, we use standard data augmentation to generate two augmented views of the same image.}}
    \label{fig-sm:derm-pairs}
\end{figure}

 Figure~\ref{fig-sm:derm-sample} shows examples of images in the Derm dataset. Figure~\ref{fig-sm:derm-pairs} shows examples of images belonging to the same patient which are taken from different viewpoints and/or from different body-parts under different lighting conditions. In the Multi Instance Contrastive Learning (MICLe) method, when multiple images of a medical condition from a given patient are available, we use two randomly selected images from all of the images that belong to this patient to directly create a positive pair of examples for contrastive learning.

\noindent \textbf{External dermatology dataset details. } The dataset used for evaluating the out-of-distribution generalization performance of the model on the dermatology task was collected by a chain of skin cancer clinics in Australia and New Zealand. When compared to the in-distribution dermatology dataset, this dataset has a much higher prevalence of skin cancers such as Melanoma, Basal Cell Carcinoma, and Actinic Keratosis. It includes 8,563 de-identified multi-image cases which we use for the purpose of evaluating the generalization of the model under distribution shift. 

\subsection{CheXpert}
\label{app:datasets-chex}
\begin{figure}[ht]
     \centering
     \includegraphics[width=1.0\textwidth]{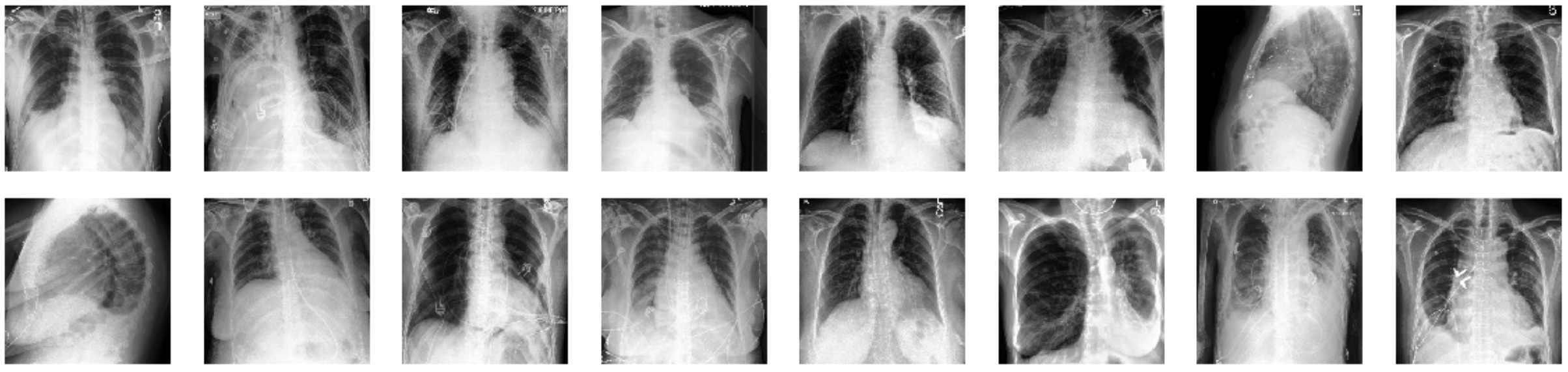}
    \caption{\small Examples images from CheXpert dataset. The chest x-rays images are less diverse in comparison to the ImageNet
and Derm dataset examples. The CheXpert task is to predict the probability of different observations from multi-view chest radiographs where we are looking for small local variations in examples using frontal and lateral radiographs.}
    \label{fig-sm:chexpert-sample}
\vspace{-.3cm}
\end{figure}

\noindent \textbf{Dataset split details.} For CheXpert dataset~\cite{irvin2019chexpert} and the task of chest X-ray interpretation, we set up the learning task to diagnose five different thoracic pathologies: atelectasis, cardiomegaly, consolidation, edema and pleural effusion. The CheXpert dataset default split contains a training set of more than 200k images and a very small validation set that contains only 200 images. This extreme size difference is mainly because the training set is constructed using an algorithmic labeler based on the free text radiology reports while the validation set is manually labeled by board-certified radiologists. Similar to Neyshabur \textit{et al.}~\cite{neyshabur2020being, raghu2019transfusion} findings, we realized due to the small size of the validation set, and the discrepancy between the label collection of the training set and the validation set, the high variance in studies is plausible. This variance implies that high performance on subsets of the training set would not correlate well with performance on the validation set, and consequently, complicating model selection from the hyper-parameter sweep. Following Neyshabur \textit{et al.}~\cite{neyshabur2020being} suggestion, in order to facilitate a robust comparison of our method to standard approaches, we define a custom subset of the training data as the validation set where we randomly re-split the full training set into 67,429 training images, 22,240 validation and 33,745 test images, respectively. This means the performances of our models are not compatible to those reported in~\cite{irvin2019chexpert} and the corresponding competition leader-board\footnote{\url{https://stanfordmlgroup.github.io/competitions/chexpert/}} for this specific dataset; nonetheless, we believe the relative performance of models is representative, informative, and comparable with~\cite{neyshabur2020being, raghu2019transfusion}. Figure~\ref{fig-sm:chexpert-sample} shows examples of images in the CheXpert dataset which includes both frontal and lateral radiographs.

\paragraph{CheXpert data augmentation.} Due to the less versatile nature of CheXpert dataset (see Fig.~\ref{fig-sm:chexpert-sample}), we used fairly strong data augmentation in order to prevent overfitting and improve final performance. At training time, the following preprocessing was applied: (1) random rotation by angle $\delta \sim U(-20, 20)$ degree, (2) random crop to 224$\times$224 pixels, (3) random left-right flip with probability 50\%, (4) linearly rescale value range from [0, 255] to [0, 1] followed by random additive brightness modulation and random multiplicative contrast modulation. Random additive brightness modulation adds a $\delta \sim U(-0.2, 0.2)$ to all channels.  Random multiplicative contrast modulation multiplies per-channel standard deviation by a factor $s \sim U(-0.2, 0.2)$. After these steps we re-clip values to the range of [0, 1].

\section{Additional Results and Experiments}
\label{app:technical}
\subsection{Dermatology Classification}

\subsubsection{Evaluation Details and Statistical Significance Testing}
\label{app:technical-stat}
To evaluate the dermatology condition classification model performance, we compared its predicted differential diagnosis with the majority voted reference standard differential diagnosis (ground-truth label) using the top-k accuracy and the average top-k sensitivity. The top-k accuracy measures how frequently the top $k$ predictions match any of the primary diagnoses in the ground truth. The top-k sensitivity measures this for each of the 26 conditions separately, whereas the final average top-k sensitivity is the average across the 26 conditions. Averaging across the 26 conditions avoids biasing towards more common conditions. We use both the top-1 and top-3 metrics in this paper.

In addition to our previous result comparing MICLe and SimCLR models against the supervised baselines, the non-parametric bootstrap is used to estimate the variability around model performance and investigating any significant improvement in the results using self-supervised pretrained models. Unlike the previous studies which uses confidence intervals obtained by multiple separate runs, for statistical significance testing, we select the best fine-tuned models for each of the architectures and compute the difference in top-1 and top-3 accuracies on bootstrap replicas of the test set.  Given predictions of two models, we generate 1,000 bootstrap replicates of the test set and computing the difference in the target performance metric (top-k accuracy and AUCs) for both models after performing this randomization. This produces a distribution for each model and we use the 95\% bootstrap percentile intervals to assess significance at the $p = 0.05$ level.

Table~\ref{tab-sm:3 } shows the comparison of the best self-supervised models \vs{} supervised pretraining on dermatology classification. Our results suggest that, MICLe models can significantly ($p<0.05$) outperform SimCLR counterpart and BiT~\cite{kolesnikov2019big} supervised model with ResNet-101 (3$\times$) architecture over top-1 and top-3 accuracies. BiT model contains additional architectural tweaks included to boost transfer performance, and was trained on a significantly larger dataset of  14M  images  labelled  with  one  or  more  of  21k classes which provides us with a strong supervised baseline \vs{} the 1M images in ImageNet.

\begin{table}[h]\centering
\caption{\small Comparison of the best self-supervised models \vs{} supervised pretraining on dermatology classification. For the significance testing, we use bootstrapping to generate the confidence intervals. Our results show that the best MICLe model can significantly outperform BiT~\cite{kolesnikov2019big} which is a very strong supervised pretraining baseline trained on ImageNet-21k.}\label{tab-sm:3 }
\vspace{-7pt}
\small
\begin{tabular}{l|l|ccc}\toprule
Architecture &Method &Top-1 Accuracy &Top-3 Accuracy \\\midrule
\multirow{2}{*}{ResNet-152 (2$\times$)} 
&MICLe ImageNet~$\!\rightarrow\!$~Derm (ours)                           &0.7037$\pm$0.0233   &0.9273$\pm$0.0133 \\
&SimCLR ImageNet~$\!\rightarrow\!$~Derm~\cite{chen2020simple}           &0.6970$\pm$0.0243   &0.9266$\pm$0.0135 \\ \midrule
\multirow{2}{*}{ResNet-50 (4$\times$)} 
&MICLe ImageNet~$\!\rightarrow\!$~Derm (ours)                           &0.7019$\pm$0.0224   &0.9247$\pm$0.0135 \\
&SimCLR ImageNet~$\!\rightarrow\!$~Derm~\cite{chen2020simple}           &0.6975$\pm$0.0240   &0.9271$\pm$0.0125 \\ \midrule
ResNet-101 (3$\times$) 
&BiT Supervised~\cite{kolesnikov2019big}              &0.6845$\pm$0.0228   &0.9143$\pm$0.0142 \\
\bottomrule
\end{tabular}
\end{table}

\subsubsection{Augmentation Selection for Multi-Instance Contrastive (MICLe) Method }
\label{app:technical-aug}
To systematically study the impact of data augmentation in our multi-instance contrastive learning framework performance, we consider two augmentation scenarios: (1) performing standard simCLR augmentation which includes random color augmentation, crops with resize, Gaussian blur, and random flips, (2) performing a partial and lightweight augmentation based on random cropping and relying only on pair selections steps to create positive pairs. To understand the importance of augmentation composition in MICLe, we pretrain models under different augmentation and investigate the performance of fine-tuned models for the dermatology classification task. As the results in Table~\ref{tab-sm:aug} suggest, MICLe under partial augmentation often outperform the full augmentation, however, the difference is not significant. We leave comprehensive investigation of the optimal augmentations to future work. 

\begin{table*}[h]\centering
\caption{\small Comparison of dermatology classification performance fine-tuned on representation learned using different unlabeled dataset with MICLe along with standard augmentation and partial augmentation. Our results suggest that MICLe under partial augmentation often outperform the full augmentation.}\label{tab-sm:aug}
\vspace{-7pt}
\footnotesize
\begin{tabular}{l|l|l|ccc}\toprule
Architecture &  Method &          Augmentation &                     Top-1 Accuracy &      Top-1 Sensitivity & AUC  \\ \midrule
\multirow{4}{*}{ResNet-152 (2$\times$)} 
& \multirow{2}{*}{MICLe Derm} 
&      Full Augmentation &                  0.6697 &                 0.5060 &         0.9562 \\
&&  Partial Augmentation &         \textbf{0.6761} &                 0.5106 &         0.9562 \\  \cmidrule{2-6}
& \multirow{2}{*}{MICLe ImageNet~$\!\rightarrow\!$~Derm} 
&      Full Augmentation &                  0.6928 &                 0.5136 &         0.9634 \\ 
&&  Partial Augmentation &                  0.6889 &                 0.5300 &         0.9620 \\
\midrule
\multirow{4}{*}{ResNet-50 (4$\times$)} 
& \multirow{2}{*}{MICLe Derm} 
&      Full Augmentation &                  0.6803 &                 0.5032 &         0.9608 \\
&&  Partial Augmentation &         \textbf{0.6808} &                 0.5204 &         0.9601 \\ \cmidrule{2-6}
& \multirow{2}{*}{MICLe ImageNet~$\!\rightarrow\!$~Derm}
&      Full Augmentation &                  0.6916 &                 0.5159 &         0.9618 \\
&&  Partial Augmentation &         \textbf{0.6938} &                 0.5087 &         0.9629 \\
\bottomrule
\end{tabular}
\end{table*}

\subsubsection{Benefits of Longer Training}

\begin{figure}[t]
     \centering
         \includegraphics[width=.24\textwidth]{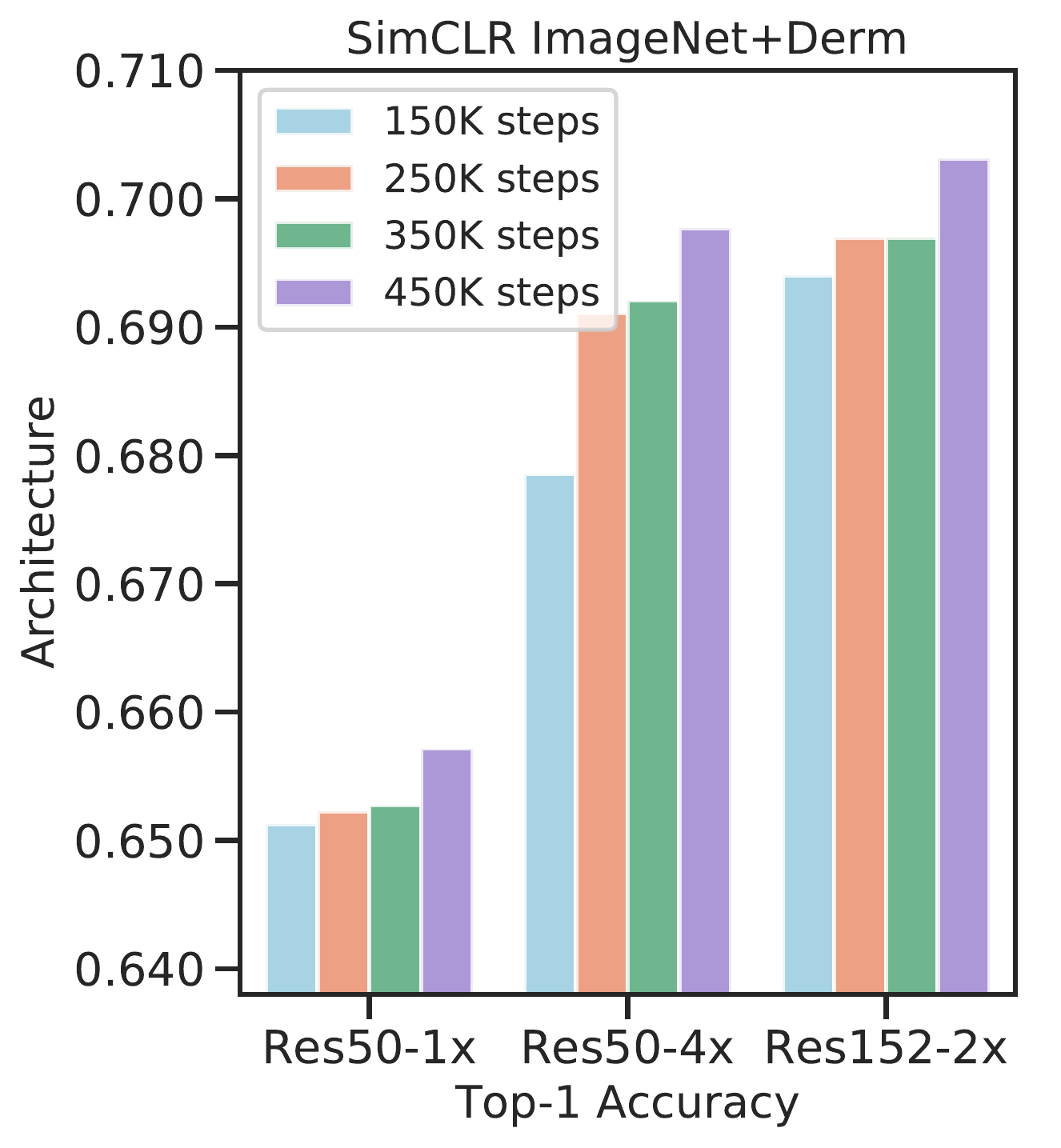}
         \includegraphics[width=.24\textwidth]{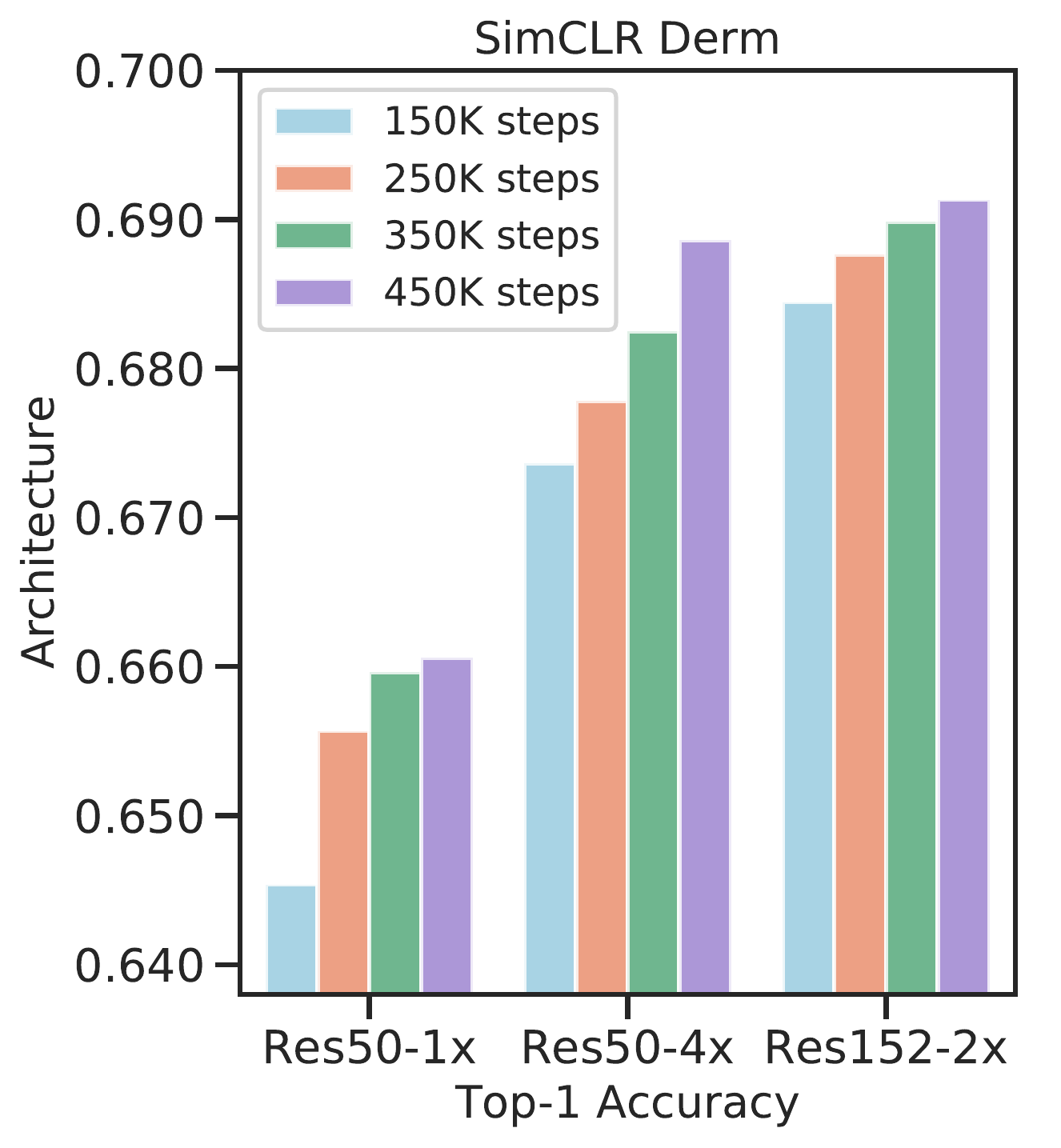}
         \includegraphics[width=.24\textwidth]{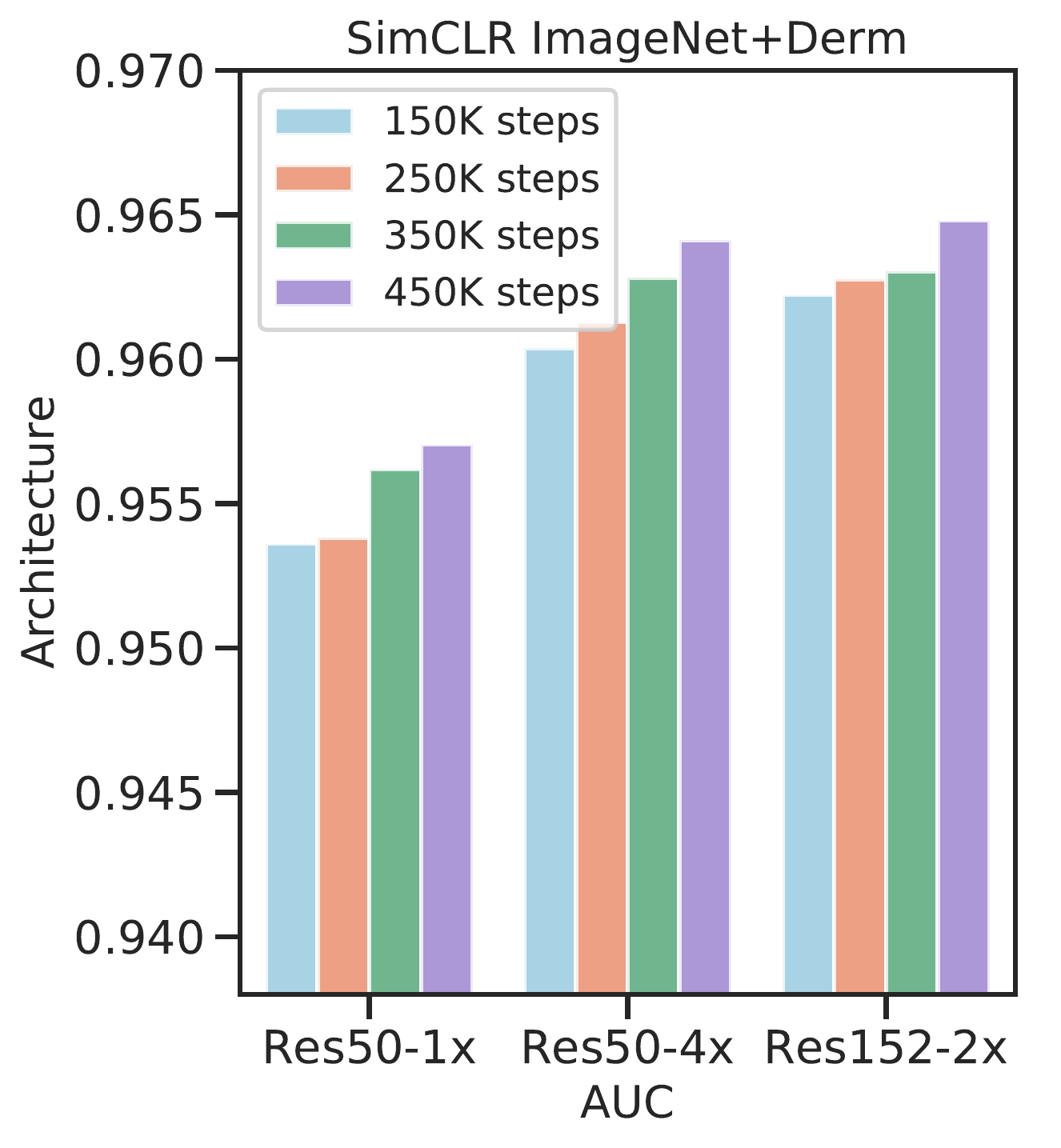}
         \includegraphics[width=.24\textwidth]{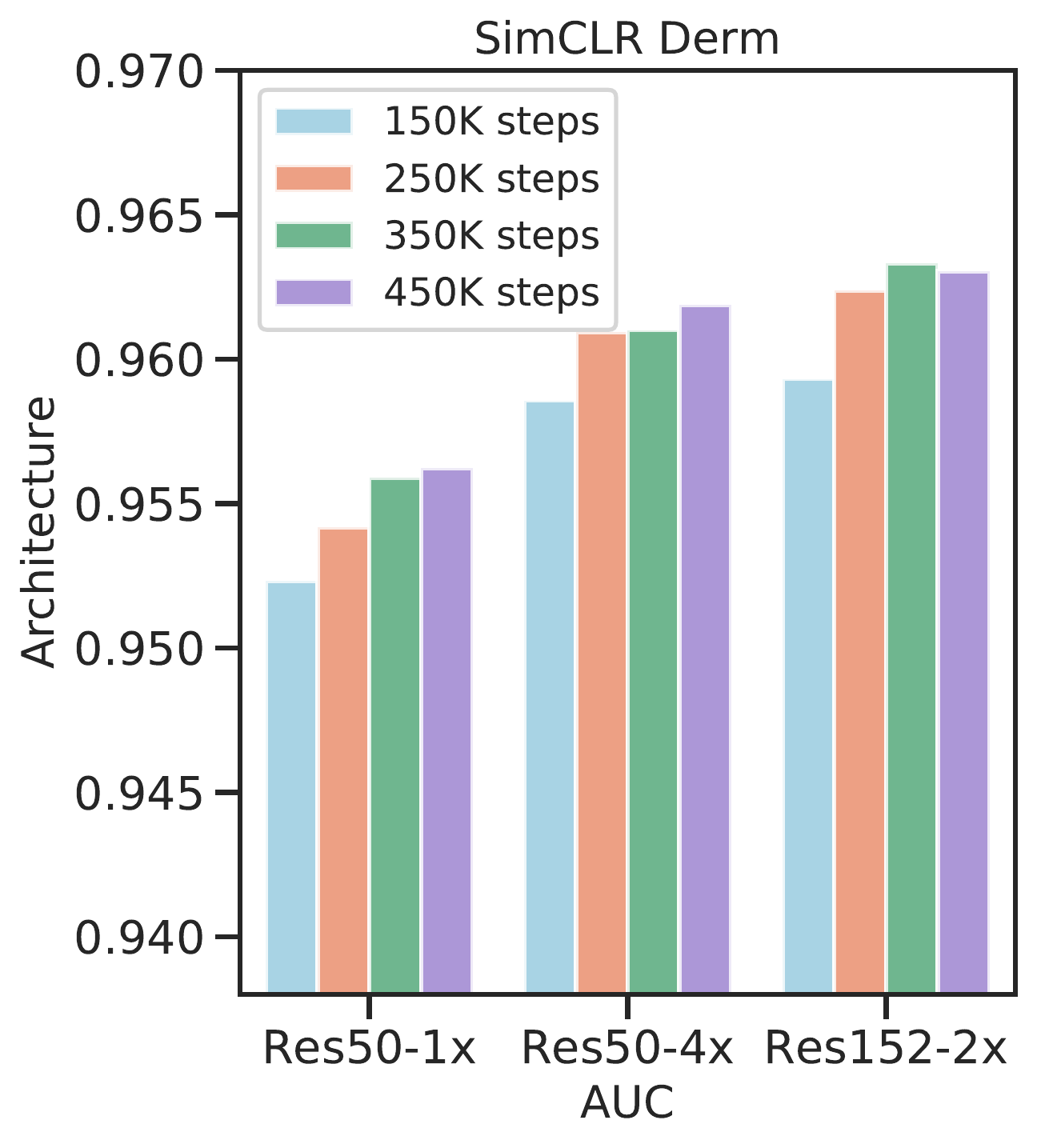}
        \vspace{-.3cm}
        \caption{\small Performance of dermatology condition classification models measured by the top-1 accuracy across different architecture and pretrained for 150,000 steps to 450,000 steps with a fixed batch size of 1024. Training longer provides more negative examples, improving the performance. Also, the results suggest that ImageNet initialization facilitating convergence, however, the performance gap between ImageNet initialized models and medical image only models are getting narrower.}
        \label{fig-sm:longer-acc}
        \vspace{0cm}
\end{figure}

Figure~\ref{fig-sm:longer-acc} shows the impact of longer training when models are pretrained for different numbers of epochs/steps. As suggested by Chen \textit{et al.}~\cite{chen2019self,chen2020big} training longer also provides more negative examples, improving the results. In this study we use a fixed batch size of 1024 and we find that with more training epochs/steps, the gaps between the performance of ImageNet initialized models with medical image only models are getting narrow, suggesting ImageNet initialization facilitating convergence where by taking fewer steps we can reach a given accuracy faster.

\begin{figure}[t]
     \centering
         \includegraphics[width=.35\textwidth]{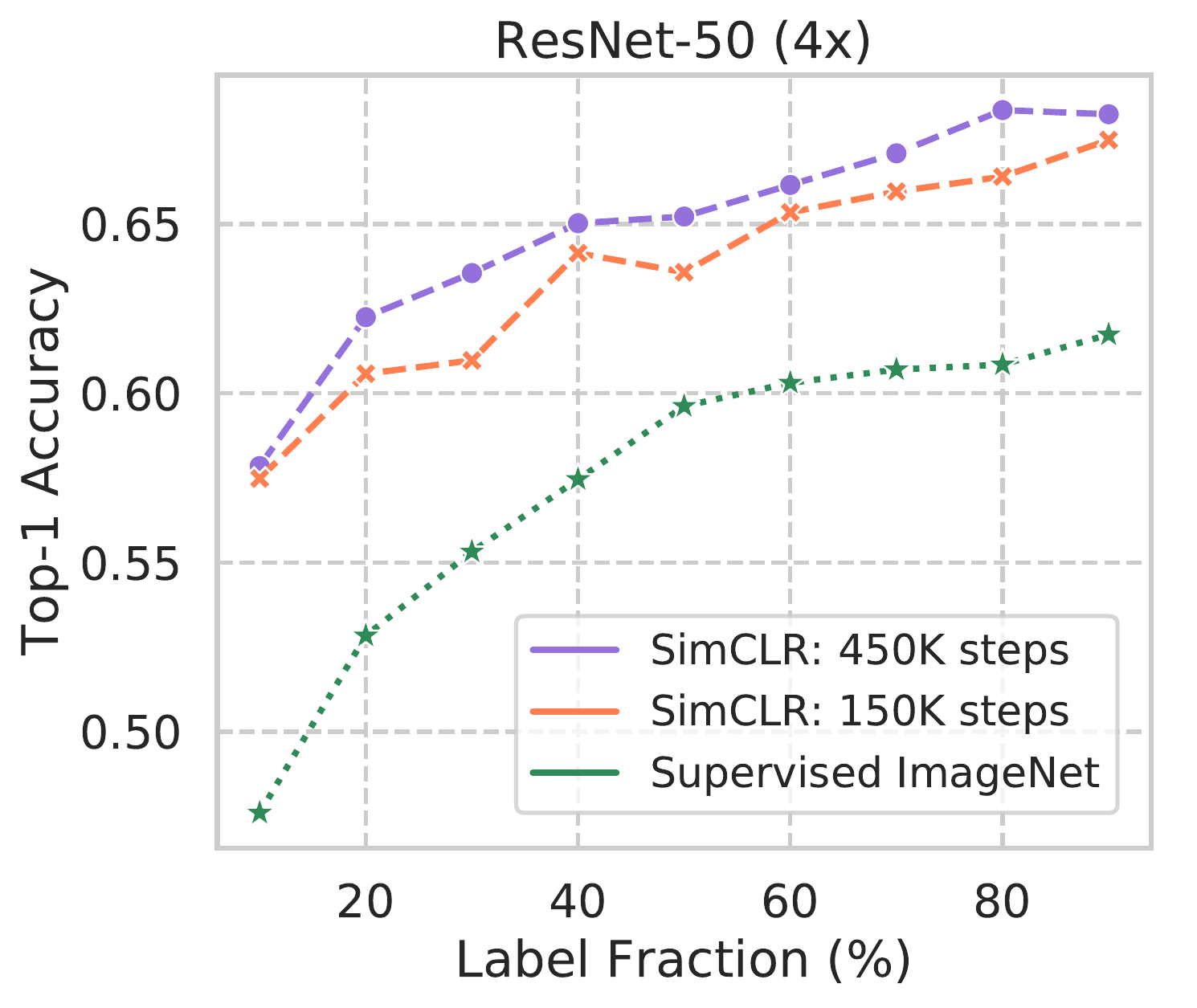}
         \includegraphics[width=.35\textwidth]{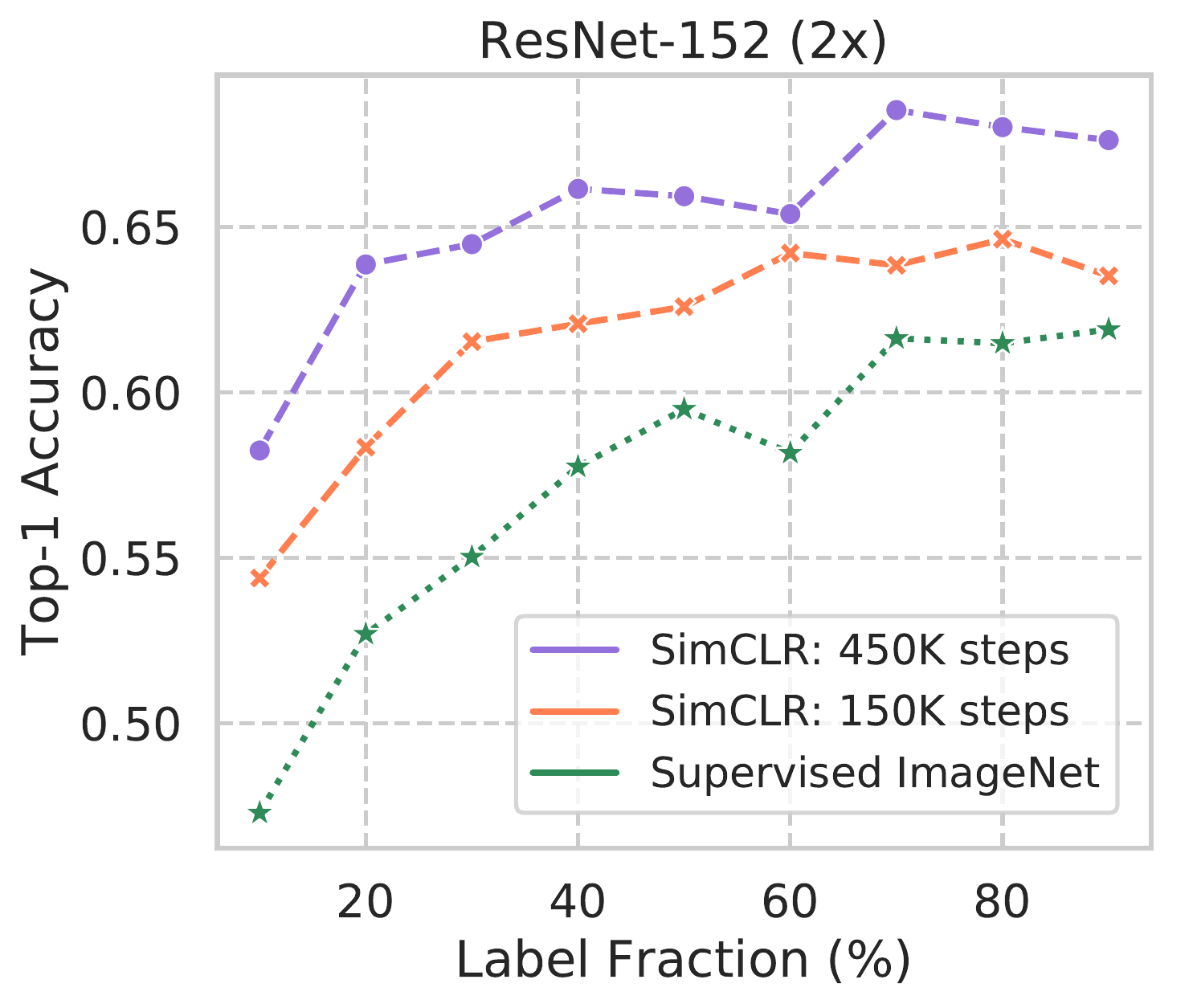}
        \vspace{-.3cm}
        \caption{\small Label efficiency progress over longer training for dermatology condition classification. The models are trained using ImageNet$\rightarrow$Derm SimCLR for 150K steps and 450K steps and fine-tuned with varied sizes of label fractions. The Supervised ImageNet used as the baseline.}
        \label{fig-sm:longer-data-eff}
        \vspace{0cm}
\end{figure}

Furthermore, Fig.~\ref{fig-sm:longer-data-eff} shows how the performance varies using the different available label fractions for dermatology task for the models pretrained for 150K steps and 450,000 steps using SimCLR ImageNet$\rightarrow$Derm dataset. These results suggest that longer training yields proportionally larger gain for different label fractions. Also, this performance gain is more pronounced in ResNet-152 (2$\times$).  In fact, for ResNet-152 (2$\times$) longer self supervised pretraining enable the model to match baseline using less than 20\% of the training data \vs{} 30\% of the training data for 150,000 steps of pretraining. 

\subsubsection{Detailed Performance Results} 
Table~\ref{tab-sm:1} shows additional results for the performance of dermatology condition classification model measured by top-1 and top-3 accuracy, and area under the curve (AUC) across different architectures. Each model is fine-tuned using transfer learning from pretrained model on ImageNet, only unlabeled medical data, or pretrained using medical data initialized from ImageNet pretrained model. Again, we observe that bigger models yield better performance across accuracy, sensitivity and AUC for this task. 

As shown in Table~\ref{tab-sm:1}, we once again observe that self-supervised pretraining with both ImageNet and in-domain Derm data is beneficial, outperforming self-supervised pretraining on ImageNet or Derm data alone. Moreover, comparing the performance of self-supervised models with Random and Supervised pretraining baseline, we observe self-supervised models significantly outperforms baselines ($p<0.05$), even using smaller models such as ResNet-50 (1$\times$). 

Table~\ref{tab-sm:2} shows additional dermatology condition classification performance for models fine-tuned on representations learned using different unlabeled datasets, and with and without multi instance contrastive learning (\multisim). Our results suggest that MICLe constantly improves the performance of skin condition classification over SimCLR~\cite{chen2019self, chen2020big}. Using statistical significance test, we observe significant improvement for top-1 accuracy using MICLe for each dataset setting ($p<0.05$).

\begin{table*}[h]\centering
\caption{\small Performance of dermatology condition classification models measured by top-1 and top-3 accuracy, and area under the curve (AUC) across different architectures. Models are pretrained for 150K steps and each model is fine-tuned using transfer learning from pretrained model on ImageNet, only unlabeled medical data, or pretrained using medical data initialized from ImageNet pretrained model. We observe that bigger models yield better performance.}\label{tab-sm:1}
\vspace{-7pt}
\footnotesize
\begin{tabular}{l|l|cc|cc|ccc}\toprule
Architecture &Method    &Top-1 Accuracy     &Top-3 Accuracy     &Top-1 Sensitivity  &Top-3 Sensitivity  &AUC \\\midrule
\multirow{5}{*}{ResNet-50 (1$\times$)} 
&SimCLR ImageNet                          &0.6258$\pm$0.0080  &0.8943$\pm$0.0041  &0.4524$\pm$0.0142  &0.7388$\pm$0.0095  &0.9480$\pm$0.0014 \\
&SimCLR Derm                              &0.6249$\pm$0.0050  &0.8967$\pm$0.0031  &0.4402$\pm$0.0093  &0.7370$\pm$0.0078  &0.9485$\pm$0.0011 \\
&SimCLR ImageNet~$\!\rightarrow\!$~Derm   &0.6344$\pm$0.0124  &0.8996$\pm$0.0080  &0.4554$\pm$0.0229  &0.7349$\pm$0.0234  &0.9511$\pm$0.0035 \\
\cmidrule{2-7}
&Supervised ImageNet                      &0.5991$\pm$0.0174  &0.8743$\pm$0.0094  &0.4215$\pm$0.0267  &0.7008$\pm$0.0225  &0.9403$\pm$0.0044 \\
&Random Initialization                    &0.5170$\pm$0.0062  &0.8136$\pm$0.0108  &0.3155$\pm$0.0152  &0.5783$\pm$0.0031  &0.9147$\pm$0.0019 \\

\midrule

\multirow{5}{*}{ResNet-50 (4$\times$)} 
&SimCLR ImageNet                         &0.6462$\pm$0.0062  &0.9082$\pm$0.0018  &0.4738$\pm$0.0055  &0.7614$\pm$0.0093  &0.9545$\pm$0.0006 \\
&SimCLR Derm                             &0.6693$\pm$0.0079  &0.9173$\pm$0.0039  &0.4954$\pm$0.0054  &0.7822$\pm$0.0012  &0.9576$\pm$0.0013 \\
&SimCLR ImageNet~$\!\rightarrow\!$~Derm  &0.6761$\pm$0.0025  &0.9176$\pm$0.0015  &0.5028$\pm$0.0091  &0.7828$\pm$0.0075  &0.9593$\pm$0.0003 \\
\cmidrule{2-7}
&Supervised ImageNet                     &0.6236$\pm$0.0032  &0.8886$\pm$0.0024  &0.4364$\pm$0.0096  &0.7216$\pm$0.0070  &0.9464$\pm$0.0005 \\
&Random Initialization                   &0.5210$\pm$0.0177  &0.8279$\pm$0.0172  &0.3330$\pm$0.0203  &0.6228$\pm$0.0314  &0.9186$\pm$0.0060 \\
\midrule

\multirow{5}{*}{ResNet-152 (2$\times$)} 
&SimCLR ImageNet                         &0.6638$\pm$0.0002  &0.9109$\pm$0.0023  &0.4993$\pm$0.0107  &0.7716$\pm$0.0039  &0.9573$\pm$0.0016 \\
&SimCLR Derm                             &0.6643$\pm$0.0051  &0.9126$\pm$0.0008  &0.5035$\pm$0.0094  &0.7808$\pm$0.0011  &0.9558$\pm$0.0006 \\
&SimCLR ImageNet~$\!\rightarrow\!$~Derm  &0.6830$\pm$0.0018  &0.9196$\pm$0.0023  &0.5156$\pm$0.0061  &0.7891$\pm$0.0058  &0.9620$\pm$0.0006 \\
\cmidrule{2-7}
&Supervised ImageNet                     &0.6336$\pm$0.0012  &0.8994$\pm$0.0022  &0.4584$\pm$0.0162  &0.7462$\pm$0.0076  &0.9506$\pm$0.0015 \\
&Random Initialization                   &0.5248$\pm$0.0121  &0.8304$\pm$0.0127  &0.3400$\pm$0.0303  &0.6310$\pm$0.0366  &0.9202$\pm$0.0055 \\

\bottomrule
\end{tabular}
\end{table*}

\begin{table}[h]\centering
\caption{\small Dermatology condition classification performance measured by top-1 accuracy, top-3 accuracy, and AUC. Models are  fine-tuned on representations learned using different unlabeled datasets, and with and without multi instance contrastive learning (MICLe). Our results suggest that MICLe constantly improves the accuracy of skin condition classification over SimCLR.}\label{tab-sm:2}
\vspace{-7pt}
\footnotesize
\begin{tabular}{l|l|cc|cc|cc}\toprule
Architecture &Method &Top-1 Accuracy &Top-3 Accuracy &Top-1 Sensitivity &Top-3 Sensitivity &AUC \\\midrule
\multirow{5}{*}{ResNet-152 (2$\times$)} 

&MICLe Derm                                  &0.6716$\pm$0.0031 &0.9132$\pm$0.0022 &0.5140$\pm$0.0093 &0.7825$\pm$0.0027 &0.9577$\pm$0.0009 \\
&SimCLR Derm                                 &0.6643$\pm$0.0051 &0.9126$\pm$0.0008 &0.5035$\pm$0.0094 &0.7808$\pm$0.0011 &0.9558$\pm$0.0006 \\
\cmidrule{2-7}
&MICLe ImageNet~$\!\rightarrow\!$~Derm       &0.6843$\pm$0.0029 &0.9246$\pm$0.0020 &0.5199$\pm$0.0108 &0.7933$\pm$0.0042 &0.9629$\pm$0.0007 \\
&SimCLR ImageNet~$\!\rightarrow\!$~Derm      &0.6830$\pm$0.0018 &0.9196$\pm$0.0023 &0.5156$\pm$0.0061 &0.7891$\pm$0.0058 &0.9620$\pm$0.0006 \\ \midrule

\multirow{5}{*}{ResNet-50 (4$\times$)} 

&MICLe Derm                                  &0.6755$\pm$0.0047 &0.9152$\pm$0.0014 &0.4900$\pm$0.0159 &0.7603$\pm$0.0092 &0.9583$\pm$0.0011 \\
&SimCLR Derm                                 &0.6693$\pm$0.0079 &0.9173$\pm$0.0039 &0.4954$\pm$0.0054 &0.7822$\pm$0.0012 &0.9576$\pm$0.0013 \\
\cmidrule{2-7}
&MICLe ImageNet~$\!\rightarrow\!$~Derm       &0.6881$\pm$0.0036 &0.9247$\pm$0.0011 &0.5106$\pm$0.0076 &0.7889$\pm$0.0091 &0.9623$\pm$0.0005 \\
&SimCLR ImageNet~$\!\rightarrow\!$~Derm      &0.6761$\pm$0.0025 &0.9176$\pm$0.0015 &0.5028$\pm$0.0091 &0.7828$\pm$0.0075 &0.9593$\pm$0.0003 \\ 
\bottomrule
\end{tabular}
\end{table}

\subsubsection{Detailed Label Efficiency Results} 
\begin{figure}[t]
     \centering
         \includegraphics[width=.32\textwidth]{images/label_fraction_r50_4_derm_top1.pdf}
         \includegraphics[width=.32\textwidth]{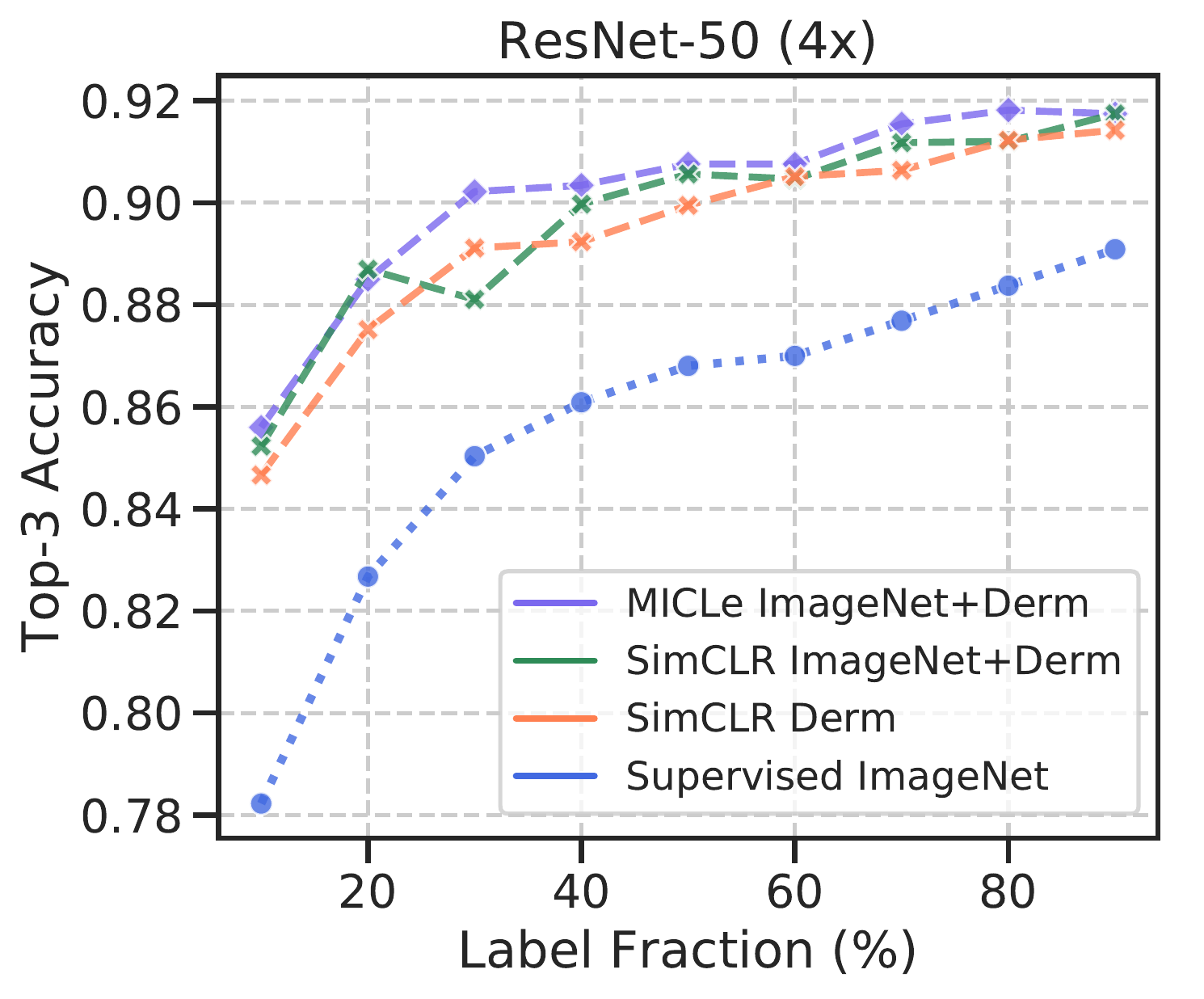}
         \includegraphics[width=.32\textwidth]{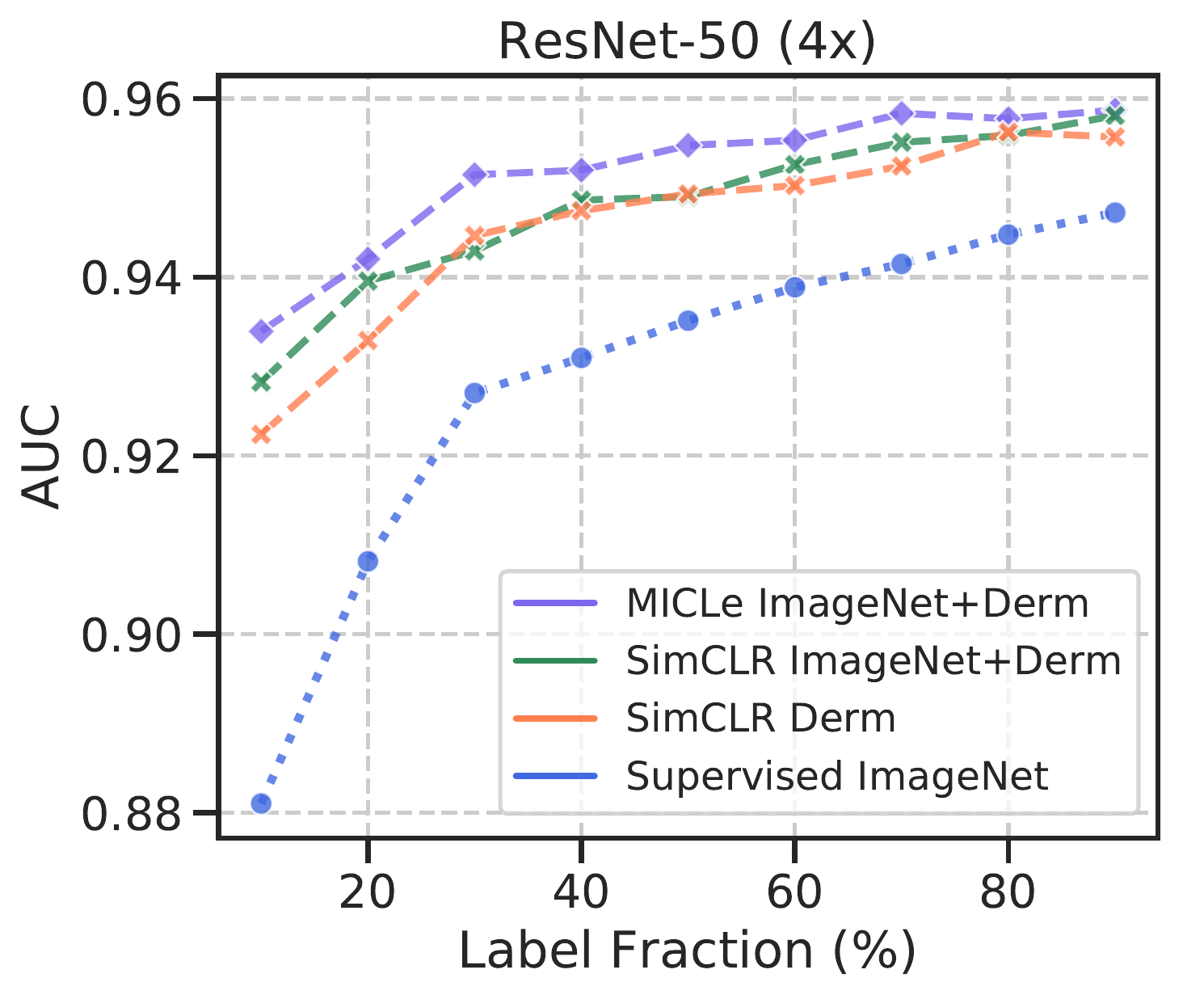}
         \\
         \includegraphics[width=.32\textwidth]{images/label_fraction_r152_2_derm_top1.pdf}
         \includegraphics[width=.32\textwidth]{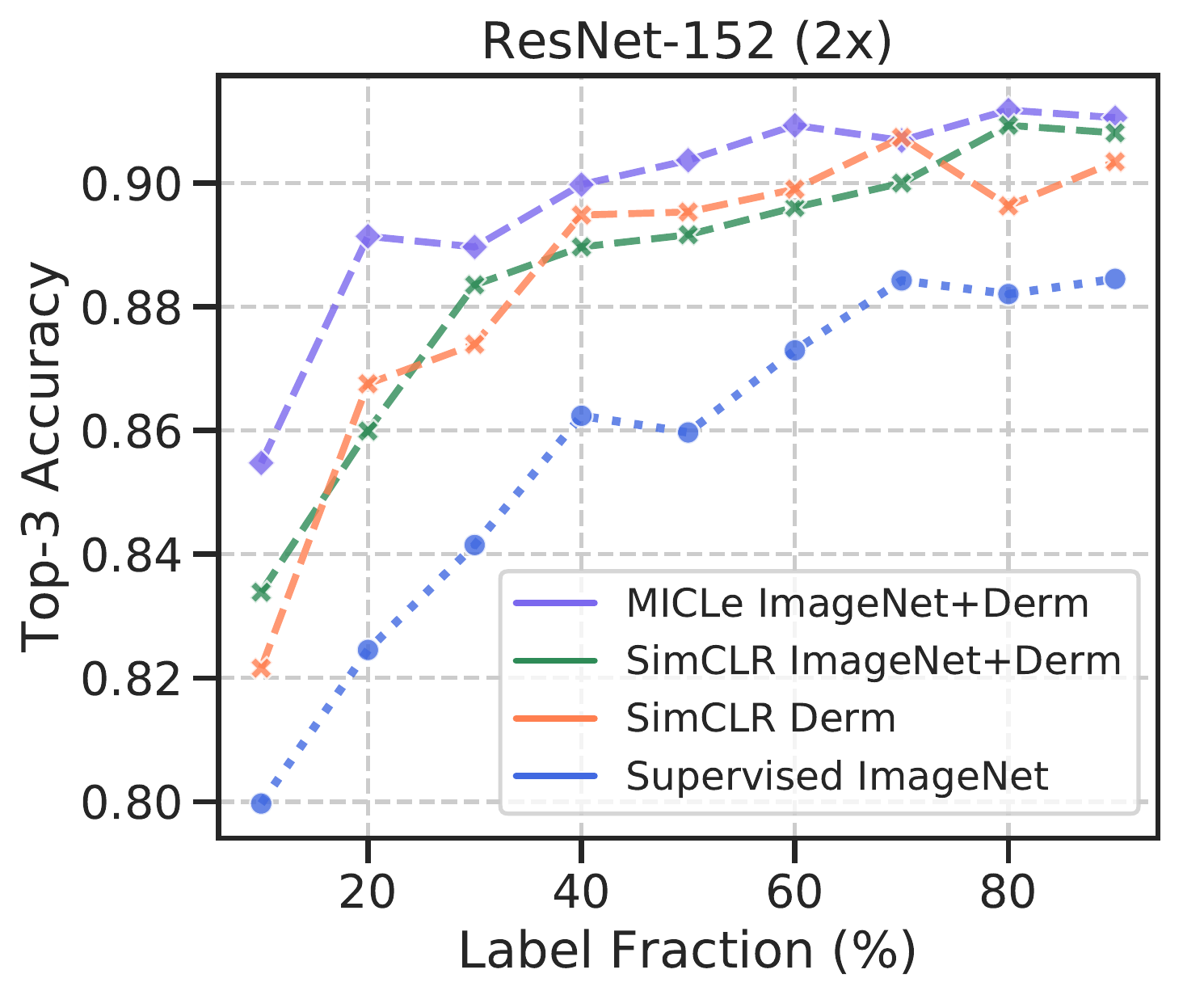}
         \includegraphics[width=.32\textwidth]{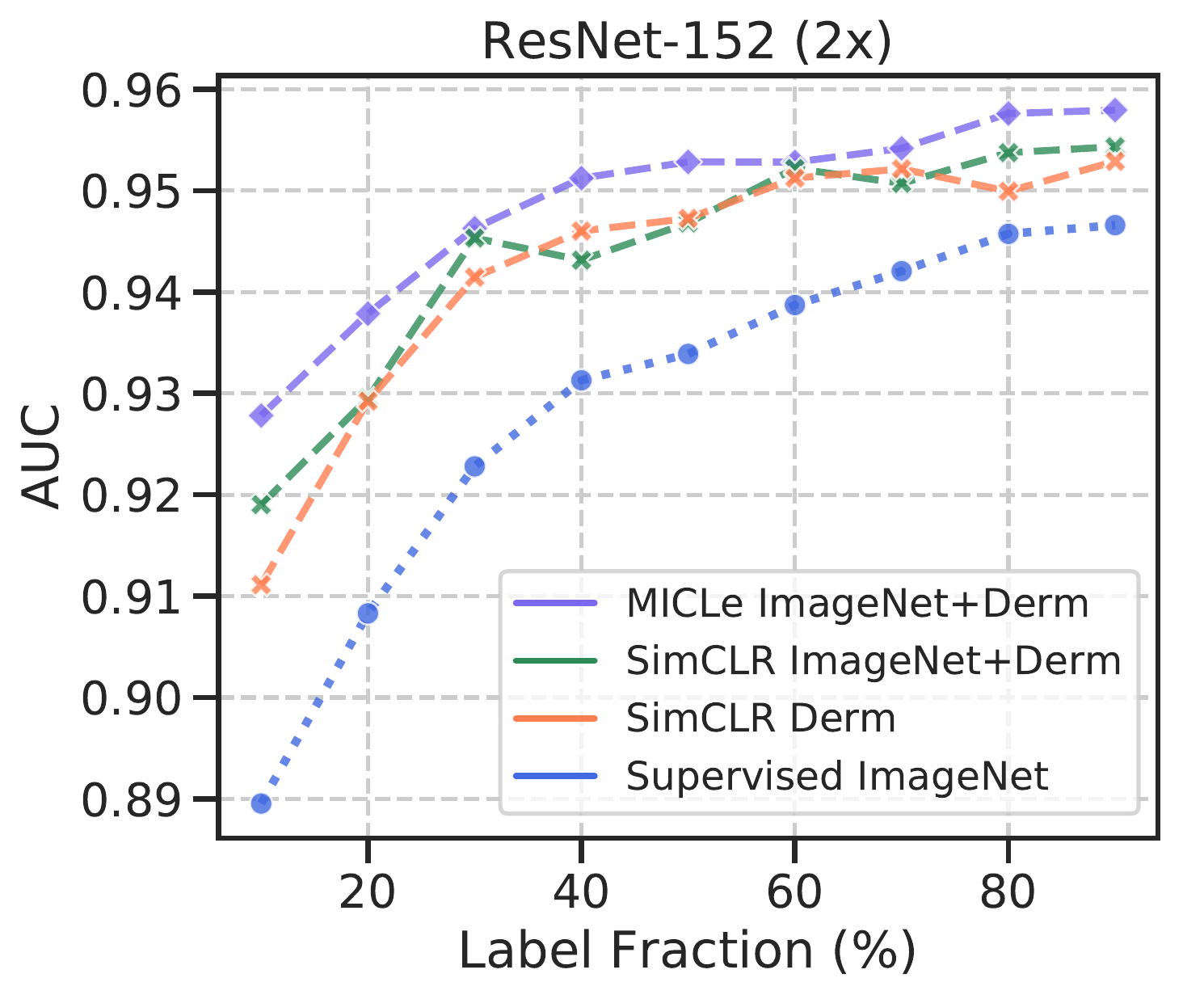}
        \vspace{-.3cm}
        \caption{\small The top-1 accuracy, top-3 accuracy, and AUC for dermatology condition classification for MICLe, SimCLR, and supervised models under different unlabeled pretraining dataset and varied sizes of label fractions. (top) ResNet-50 (4$\times$), (bottom) ResNet-152 (2$\times$). }
        \label{fig-sm:4}
        \vspace{-.1cm}
\end{figure}

Figure~\ref{fig-sm:4} and Table~\ref{tab-sm:5} provide additional performance results to investigate label-efficiency of the selected self-supervised models in the dermatology task. These results, back-up our finding that the pretraining using self-supervised models can significantly help with label efficiency for medical image classification, and in all of the fractions, self-supervised models outperform the supervised baseline. Also, we observe that MICLe yields proportionally larger gains when fine-tuning with fewer labeled examples and this is consistent across top-1 and top-3 accuracy and sensitivity, and AUCs for the dermatology classification task.

\begin{table}[h]\centering
\caption{\small Classification accuracy and sensitivity for dermatology condition classification task, obtained by fine-tuning the SimCLR and MICLe on 10\%, 50\%, and 90\%  of the labeled data. As a reference, ResNet-50 (4$\times$) fine-tuned the supervised ImageNet model and using 100\% labels achieves 62.36\% top-1 and 88.86\% top-3 accuracy.}\label{tab-sm:5}
\vspace{-7pt}
\scriptsize
\begin{tabular}{@{\hspace{.1cm}}l@{\hspace{.1cm}}|@{\hspace{.1cm}}l|@{\hspace{.1cm}}ccc@{\hspace{.1cm}}|@{\hspace{.1cm}}ccc@{\hspace{.1cm}}@{\hspace{.1cm}}|@{\hspace{.1cm}}ccc@{\hspace{.1cm}}|@{\hspace{.1cm}}ccc}\toprule
\multicolumn{2}{c|@{\hspace{.1cm}}}{Performance Metric} &\multicolumn{3}{c|@{\hspace{.1cm}}}{Top-1 Accuracy} &\multicolumn{3}{c|@{\hspace{.1cm}}}{Top-3 Accuracy} &\multicolumn{3}{c|@{\hspace{.1cm}}}{Top-1 Sensitivity} &\multicolumn{3}{c@{\hspace{.1cm}}}{Top-3 Sensitivity} \\\midrule
Architecture &Method &10\% &50\% &90\% &10\% &50\% &90\% &10\% &50\% &90\% &10\% &50\% &90\% \\ \midrule
\multirow{4}{*}{ResNet-152 (2$\times$)} 
&MICLe ImageNet~$\!\rightarrow\!$~Derm        &0.5802 &0.6542 &0.6631 &0.8548 &0.9037 &0.9105 &0.3839 &0.4795 &0.4947 &0.6496 &0.7567 &0.7720 \\
&SimCLR ImageNet~$\!\rightarrow\!$~Derm       &0.5439 &0.6260 &0.6353 &0.8339 &0.8916 &0.9081 &0.3446 &0.4491 &0.4786 &0.6243 &0.7269 &0.7792 \\
&SimCLR Derm                                  &0.5313 &0.6296 &0.6522 &0.8216 &0.8953 &0.9034 &0.3201 &0.4710 &0.4906 &0.6036 &0.7373 &0.7557 \\ \cmidrule{2-14} 
&Supervised ImageNet                          &0.4728 &0.5950 &0.6191 &0.7997 &0.8597 &0.8845 &0.2495 &0.4303 &0.4677 &0.5452 &0.7015 &0.7326 \\ \midrule
\multirow{4}{*}{ResNet-50 (4$\times$)} 
&MICLe ImageNet~$\!\rightarrow\!$~Derm        &0.5884 &0.6498 &0.6712 &0.8560 &0.9076 &0.9174 &0.3841 &0.4878 &0.5120 &0.6555 &0.7554 &0.7771 \\
&SimCLR ImageNet~$\!\rightarrow\!$~Derm       &0.5748 &0.6358 &0.6749 &0.8523 &0.9056 &0.9174 &0.3983 &0.4889 &0.5285 &0.6585 &0.7691 &0.7902 \\
&SimCLR Derm                                  &0.5574 &0.6331 &0.6483 &0.8466 &0.8995 &0.9142 &0.3307 &0.4387 &0.4675 &0.6233 &0.7412 &0.7728 \\ \cmidrule{2-14}
&Supervised ImageNet                          &0.4760 &0.5962 &0.6174 &0.7823 &0.8680 &0.8909 &0.2529 &0.4247 &0.4677 &0.5272 &0.6925 &0.7379 \\
\bottomrule
\end{tabular}
\end{table}

\subsubsection{Subgroup Analysis}

\begin{figure}[t]
     \centering
         \includegraphics[width=.40\textwidth]{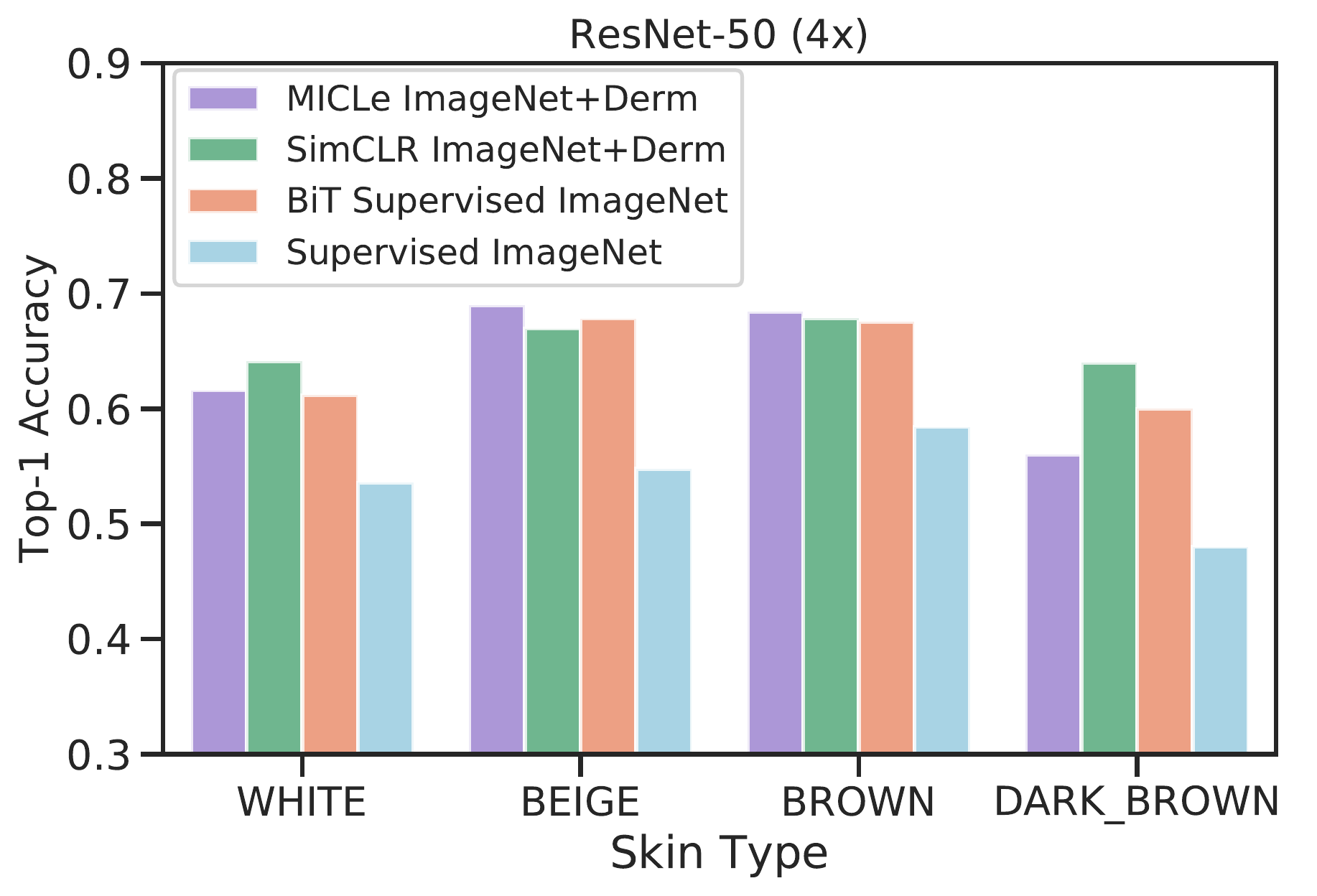} ~~~~~~
         \includegraphics[width=.40\textwidth]{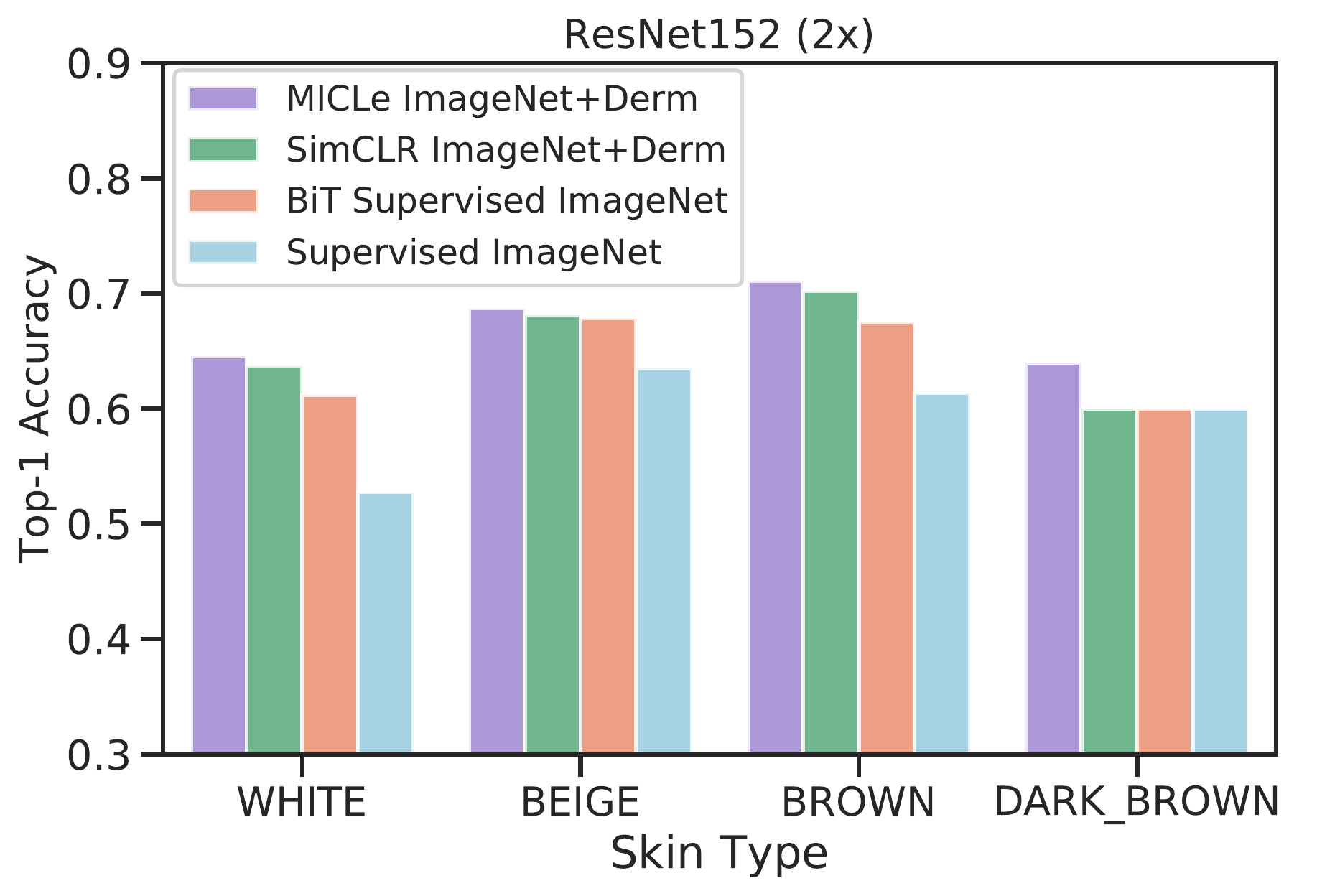} 
         \includegraphics[width=.40\textwidth]{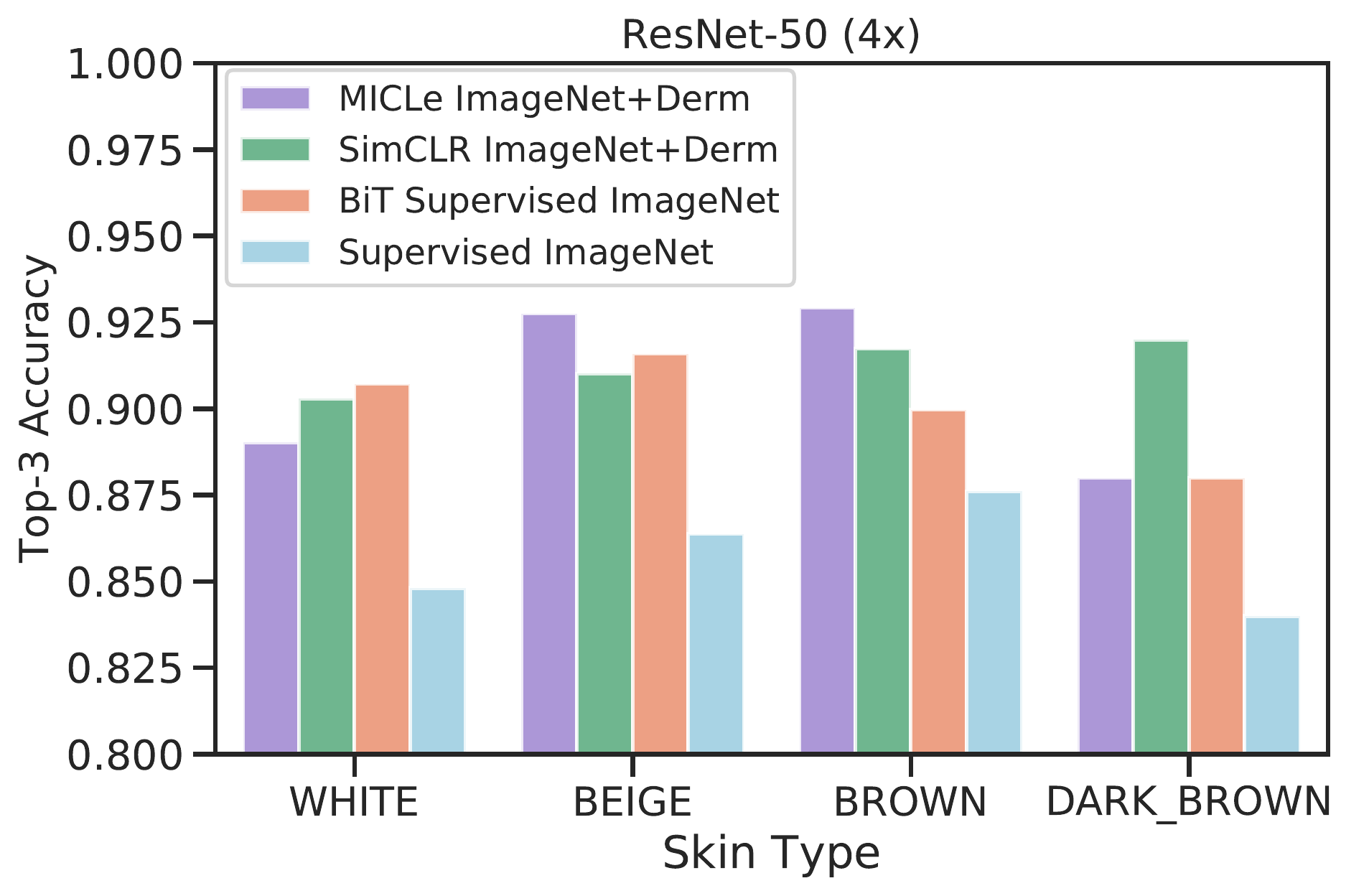}  ~~~~~~  
         \includegraphics[width=.40\textwidth]{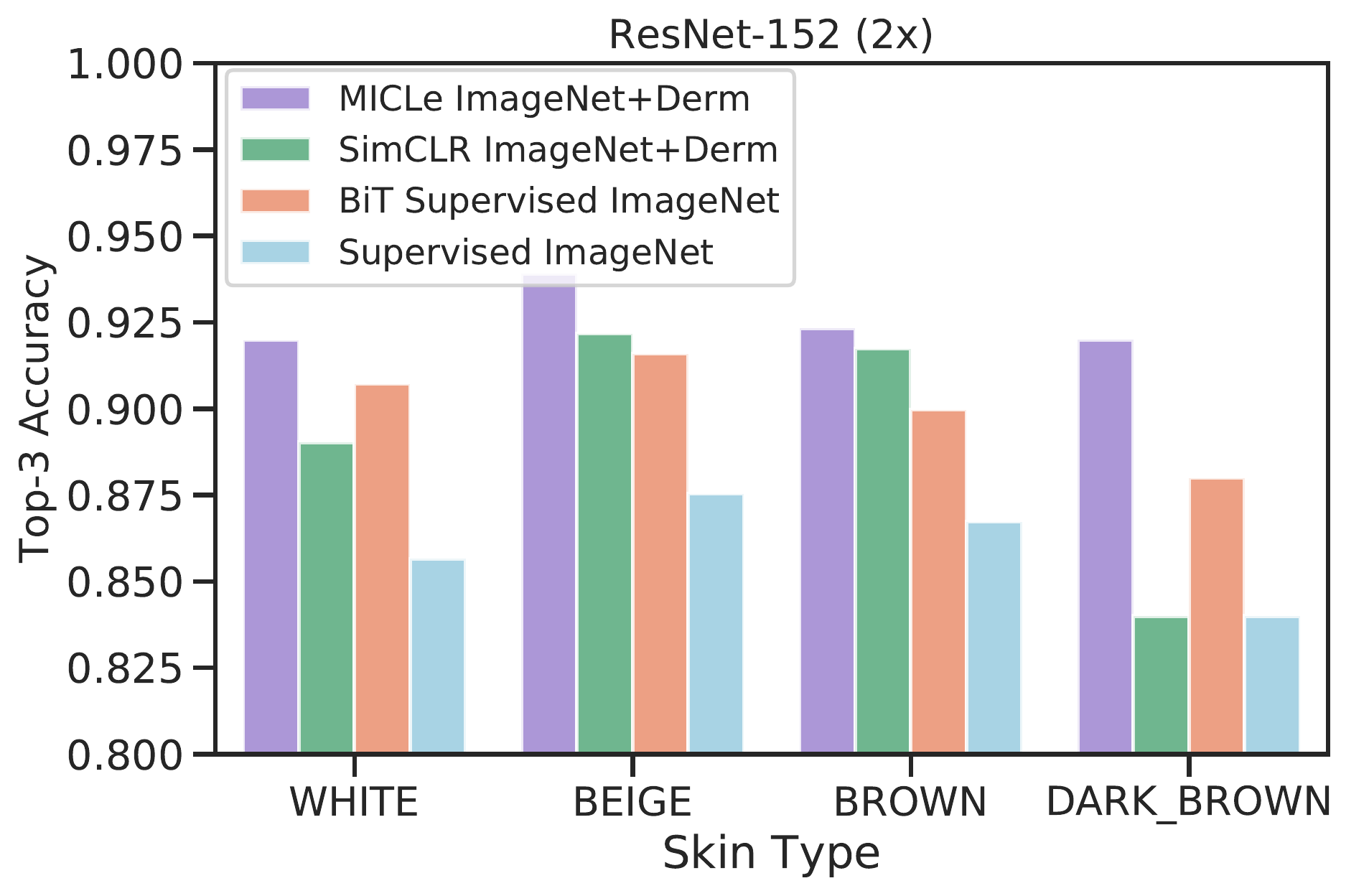} 
        \vspace{-.3cm}
        \caption{\small Performance of the different models across different skin type subgroups for the dermatology classification task. Models pretrained using self-supervised learning perform much better on the rare skin type subgroups.}
        \label{fig-sm:8}
        \vspace{-.3cm}
\end{figure}

In another experiment, we also investigated whether the performance gains when using pretrained representations from self-supervised learning are evenly distributed across different subgroups of interest for the dermatology task; it is important for deployment in clinical settings that model performance is similar across such subgroups.  
We specifically explore top-1 and top-3 accuracy across different skin types of white, beige, brown, and dark brown. Figure~\ref{fig-sm:8} shows the distribution of performance across these subgroups. We observe that while the baseline supervised pretrained model performance drops on the rarer skin types, using self-supervised pretraining, the model performance is more even across the different skin types. This exploratory experiment suggests that the learnt representations are likely general and not picking up any spurious correlations during pretraining.

\begin{figure}[t]
     \centering
         \includegraphics[width=.32\textwidth]{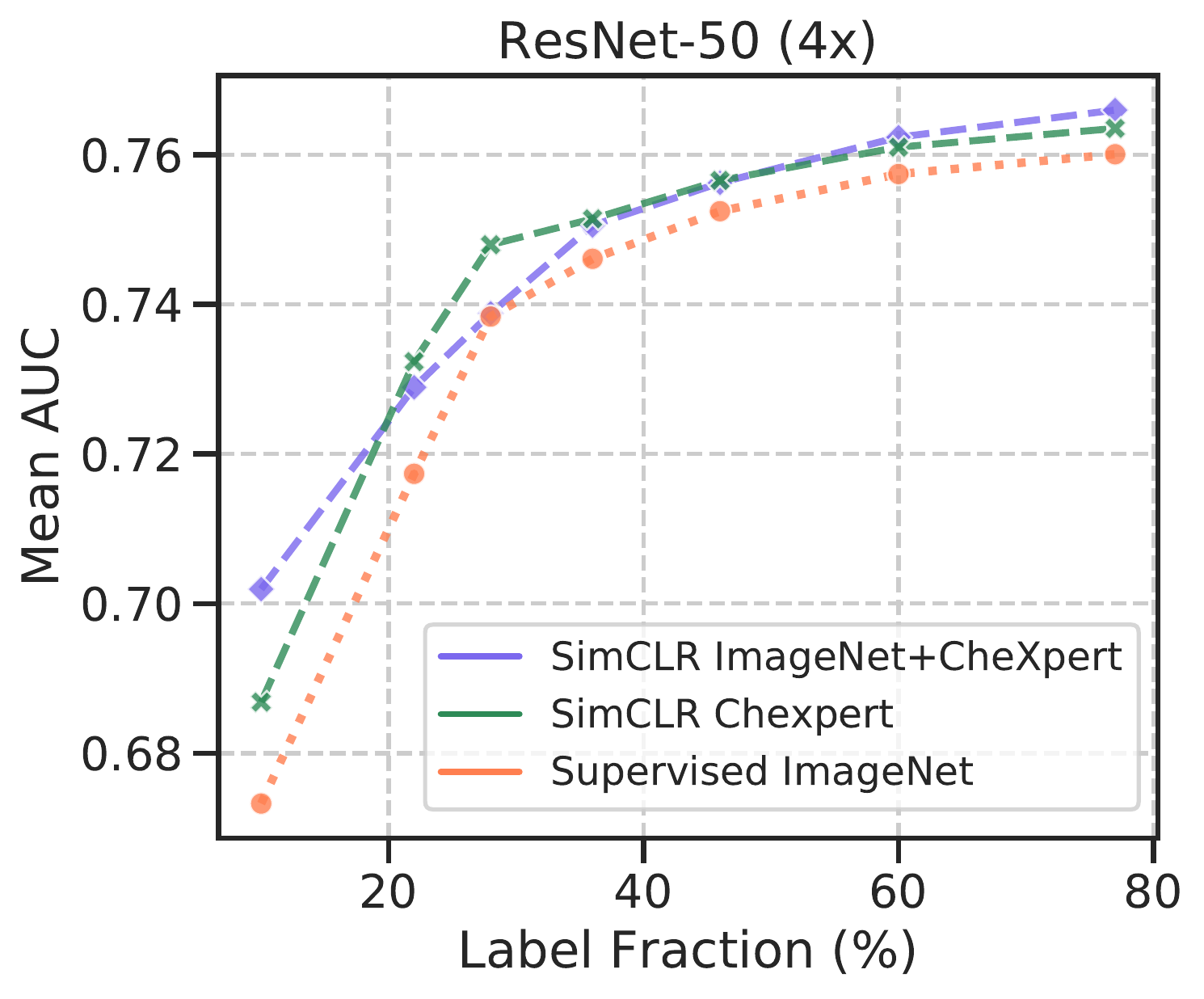}
         \includegraphics[width=.32\textwidth]{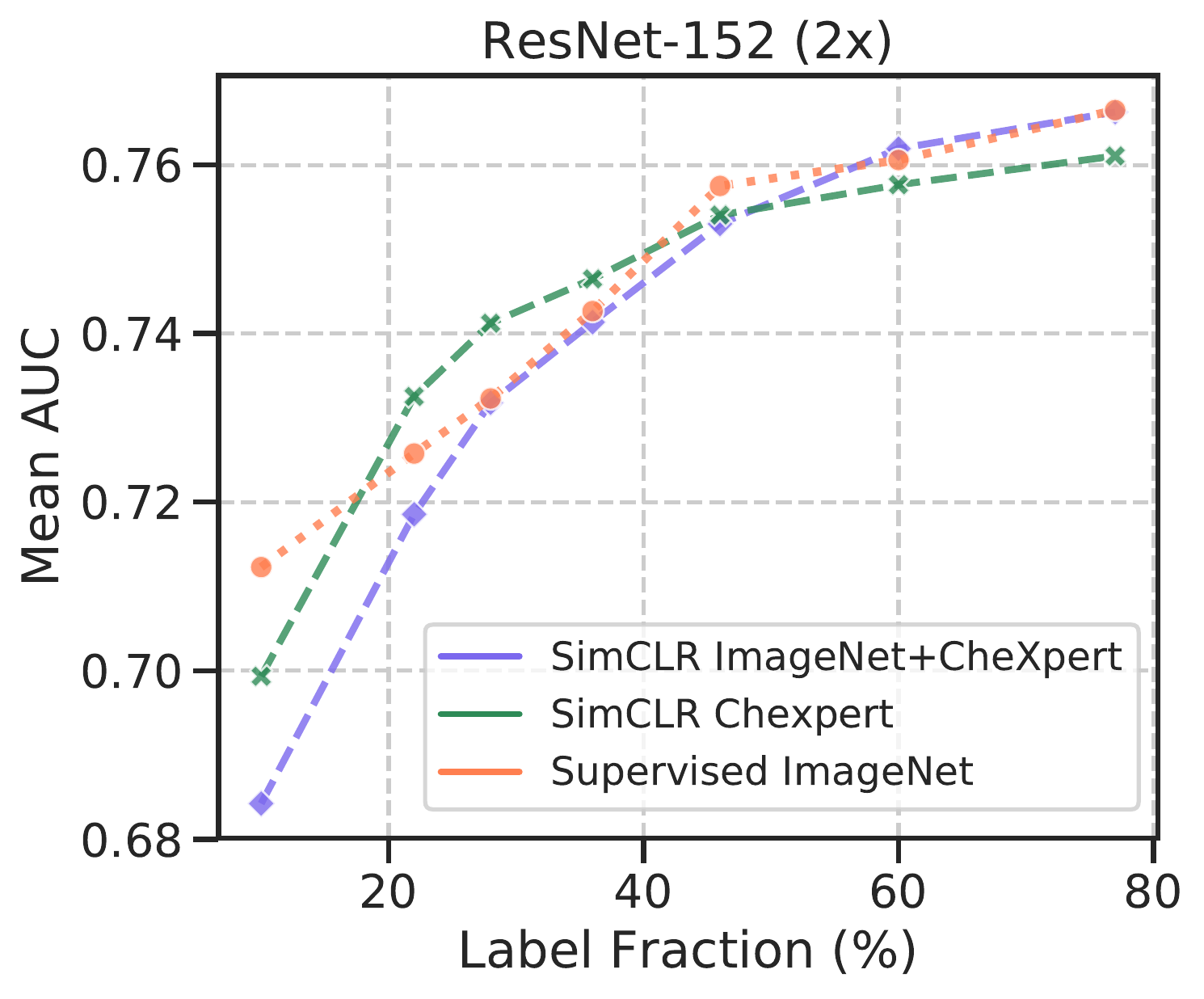}
        \vspace{-.3cm}
        \caption{\small Mean AUC for chest X-ray classification using self-supervised, and supervised pretrained models over varied sizes of label fractions for ResNet-50 (4$\times$) and ResNet-152 (2$\times$) architecture. }
        \label{fig-sm:5}
        \vspace{-.3cm}
\end{figure}

\begin{figure}[t]
     \centering
         \includegraphics[width=.32\textwidth]{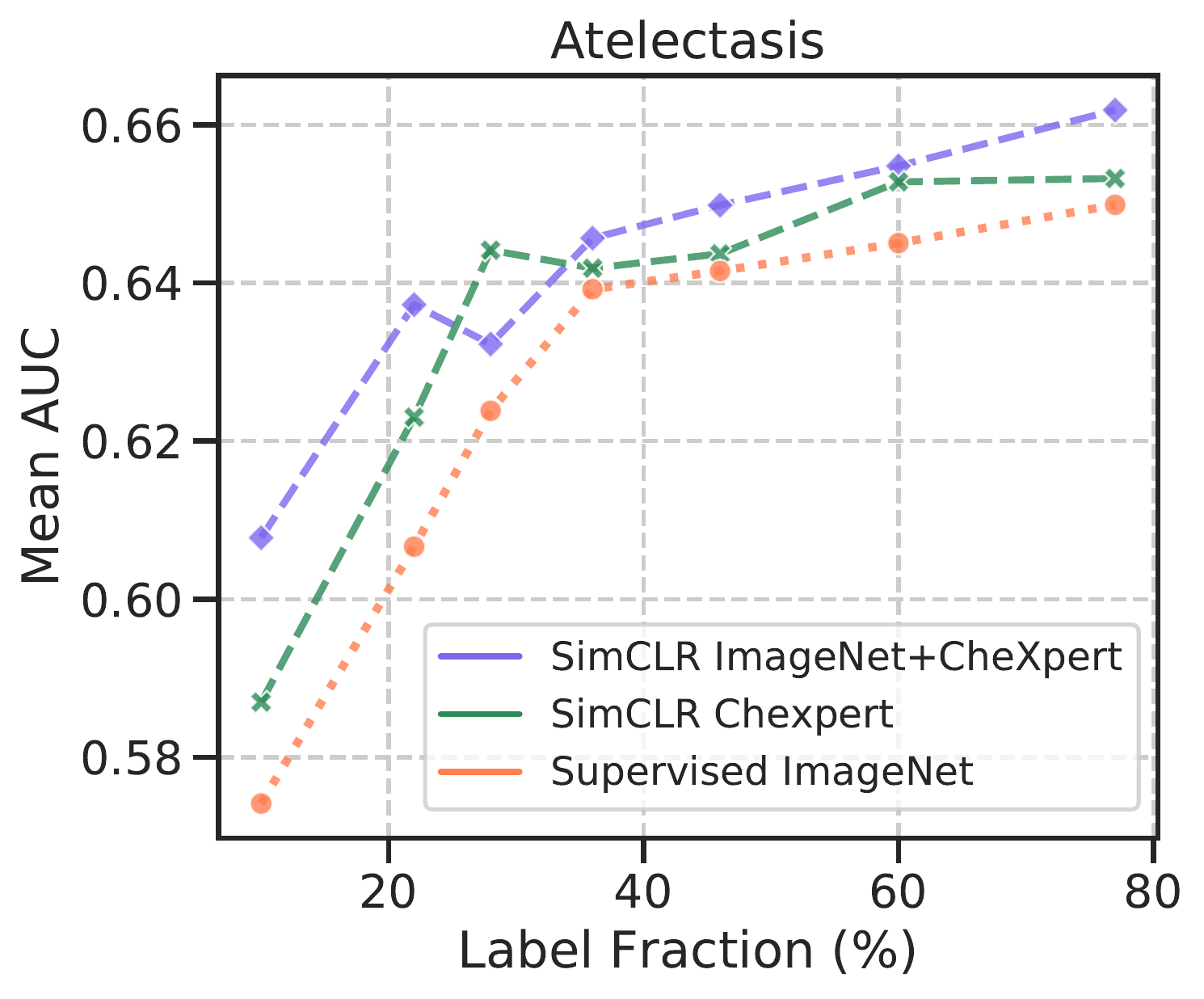} 
         \includegraphics[width=.32\textwidth]{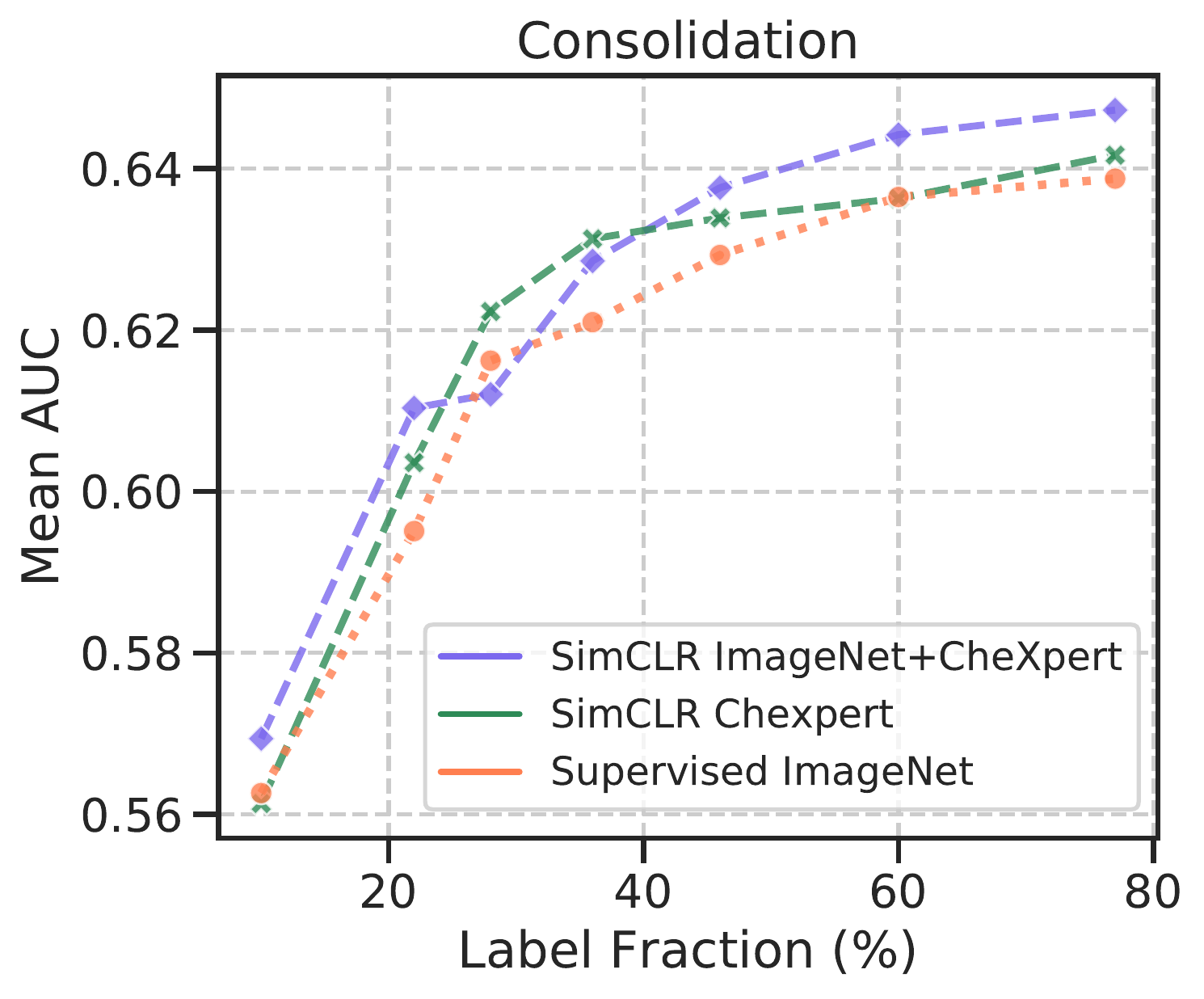}
         \includegraphics[width=.32\textwidth]{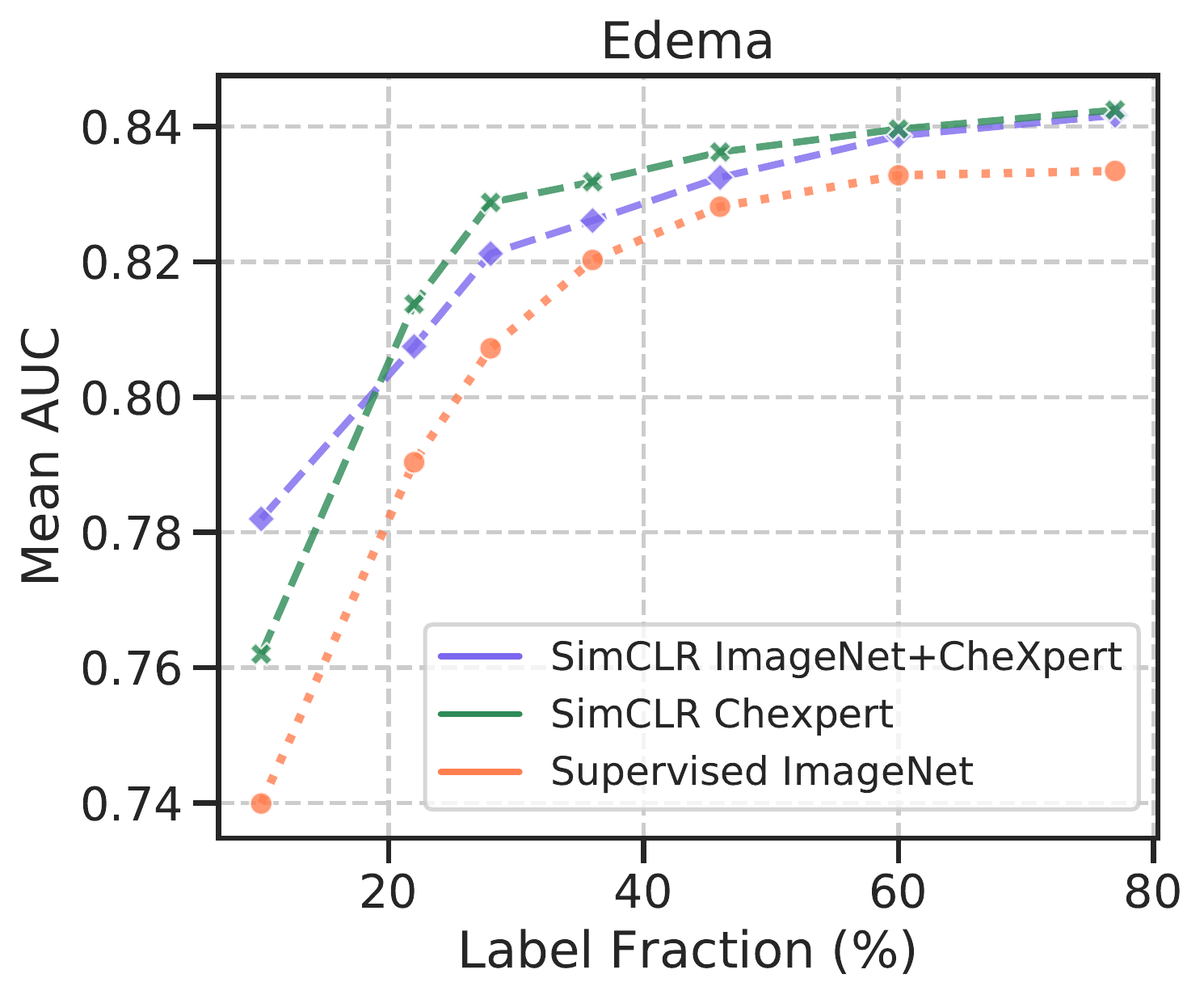} 
        \vspace{-.3cm}
        \caption{\small Performances of diagnosing different pathologies on the CheXpert dataset measured with AUC over varied sizes of label fractions for ResNet-50 (4$\times$).}
        \label{fig-sm:6}
        \vspace{-.3cm}
\end{figure}

\begin{table}[h]\centering
\caption{\small Performances of diagnosing different pathologies on the CheXpert dataset measured with AUC. The distribution of AUC performance across different pathologies suggests transfer learning, using both self-supervised and supervised models, provides mixed performance gains on this specific dataset.}\label{tab-sm:4}
\vspace{-7pt}
\footnotesize
\begin{tabular}{l|l|ccccc} \toprule
Architecture  &              Method  &        Atelectasis &        Cardiomegaly &      Consolidation &              Edema &   Pleural Effusion \\
\midrule
\multirow{4}{*}{ResNet-50 (1$\times$)}
&SimCLR ImageNet~$\!\rightarrow\!$~CheXpert          &  0.6561$\pm$0.0052 &  0.8237$\pm$0.0024 &  0.6516$\pm$0.0051 &  0.8462$\pm$0.0008 &  0.8614$\pm$0.0016 \\
&SimCLR CheXpert                                     &  0.6546$\pm$0.0030 &  0.8206$\pm$0.0025 &  0.6521$\pm$0.0027 &  0.8443$\pm$0.0012 &  0.8620$\pm$0.0005 \\
&SimCLR ImageNet                                     &  0.6516$\pm$0.0046 &  0.8190$\pm$0.0015 &  0.6456$\pm$0.0036 &  0.8431$\pm$0.0012 &  0.8610$\pm$0.0010 \\ \cmidrule{2-7}
&Supervised ImageNet                                 &  0.6555$\pm$0.0027 &  0.8188$\pm$0.0023 &  0.6517$\pm$0.0043 &  0.8429$\pm$0.0011 &  0.8607$\pm$0.0011 \\ \midrule
\multirow{4}{*}{ResNet-50 (4$\times$)}
&SimCLR ImageNet~$\!\rightarrow\!$~CheXpert          &  0.6679$\pm$0.0022 &  0.8262$\pm$0.0026 &  0.6576$\pm$0.0039 &  0.8444$\pm$0.0012 &  0.8599$\pm$0.0018 \\
&SimCLR CheXpert                                     &  0.6620$\pm$0.0038 &  0.8244$\pm$0.0017 &  0.6491$\pm$0.0029 &  0.8438$\pm$0.0014 &  0.8592$\pm$0.0013 \\
&SimCLR ImageNet                                     &  0.6633$\pm$0.0025 &  0.8228$\pm$0.0014 &  0.6525$\pm$0.0028 &  0.8439$\pm$0.0015 &  0.8641$\pm$0.0013 \\ \cmidrule{2-7}
&Supervised ImageNet                                 &  0.6570$\pm$0.0051 &  0.8218$\pm$0.0017 &  0.6546$\pm$0.0040 &  0.8425$\pm$0.0008 &  0.8624$\pm$0.0013 \\ \midrule
\multirow{4}{*}{ResNet-152 (2$\times$)}
&SimCLR ImageNet~$\!\rightarrow\!$~CheXpert          &  0.6666$\pm$0.0027 &  0.8290$\pm$0.0019 &  0.6516$\pm$0.0024 &  0.8461$\pm$0.0016 &  0.8584$\pm$0.0015 \\
&SimCLR CheXpert                                     &  0.6675$\pm$0.0040 &  0.8278$\pm$0.0015 &  0.6521$\pm$0.0030 &  0.8444$\pm$0.0013 &  0.8602$\pm$0.0016 \\
&SimCLR ImageNet                                     &  0.6621$\pm$0.0067 &  0.8239$\pm$0.0014 &  0.6495$\pm$0.0046 &  0.8439$\pm$0.0013 &  0.8637$\pm$0.0014 \\ \cmidrule{2-7}
&Supervised ImageNet                                 &  0.6496$\pm$0.0030 &  0.8224$\pm$0.0022 &  0.6498$\pm$0.0040 &  0.8408$\pm$0.0014 &  0.8615$\pm$0.0010 \\    
\bottomrule
\end{tabular}
\end{table}

\subsection{Chest X-ray Classification}
\label{app:technical-chex}
\subsubsection{Detailed Performance Results} 
For the task of X-ray interpretation on the CheXpert dataset, we set up the learning task to detect 5 different pathologies: atelectasis, cardiomegaly, consolidation, edema and pleural effusion. Table~\ref{tab-sm:4} shows the AUC performance on the different pathologies on the CheXpert dataset. We once again observe that self-supervised pretraining with both ImageNet and in-domain medical data is beneficial, outperforming self-supervised pretraining on ImageNet or CheXpert alone. Also, the distribution of AUC performance across different pathologies suggests transfer learning, using both self-supervised and supervised models, provides mixed performance gains on this specific dataset. These observations are aligned with the findings of~\cite{raghu2019transfusion}. Although less pronounced, once again we observe that bigger models yield better performance.

\subsubsection{Detailed Label-efficiency Results}
Figure~\ref{fig-sm:5} and Fig.~\ref{fig-sm:6} show how the performance changes when using different label fractions for the chest X-ray classification task. For architecture ResNet-50 (4$\times$) self supervised models consistently outperform the supervised baseline, however, this trend is less striking for ResNet-152 (2$\times$) models. We also observe that performance improvement in label efficiency is less pronounced for chest X-ray classification task in comparison to dermatology classification. We believe that with additional in-domain unlabeled data (we only use the CheXpert dataset for pretraining), self-supervised pretraining for chest X-ray classification improves.

\end{document}